\documentclass[journal]{IEEEtran}

\usepackage[utf8]{inputenc}

\usepackage{comment,rotating}

\usepackage{url}  %

\usepackage{caption,subcaption} %

\usepackage[hidelinks, breaklinks]{hyperref}

\usepackage[framemethod=TikZ]{mdframed}
\mdfdefinestyle{MyFrame}{%
    roundcorner=12pt,
    innerrightmargin=10pt,
    innerleftmargin=10pt,
    }

\usepackage{pifont}
\usepackage{hyperref}

\newcommand{\xm}{\ding{55}}%

\usepackage{diagbox} 

\usepackage{multirow}
\usepackage{xcolor,soul} 
\usepackage{multicol}
\usepackage{svg}

\usepackage{longtable}

\usepackage{graphicx}

\usepackage{pgfplots}
\usepackage{pgfplotstable}
\pgfplotsset{compat=1.7}
\usepgfplotslibrary{groupplots}
\usepgfplotslibrary{units}
\usepackage{pgf-pie}

\pgfplotsset{
SmallBarPlot/.style={
    font=\footnotesize,
    ybar,
    width=\linewidth,
    ymin=0,
    xtick=data,
    xticklabel style={text width=1.5cm, rotate=90, align=center}
},
BlueBars/.style={
    fill=blue!20, bar width=0.25
},
RedBars/.style={
    fill=red!20, bar width=0.25
}
}

\usepackage{xspace}
\newcommand{\attack}{ATT\&CK\xspace}
\newcommand{\dettect}{DeTT\&CT\xspace}

\usepackage[nolist]{acronym}

\usepackage[textsize=scriptsize,textwidth=3em]{todonotes}

\usepackage[normalem]{ulem}

\newif\ifNotes
\Notesfalse

\newif\ifExamples
\Examplestrue

\begin{document}

\title{SoK: The MITRE \attack Framework in Research and Practice}

\author{Shanto Roy,
Emmanouil Panaousis,
Cameron Noakes,
Aron Laszka, 
Sakshyam Panda, and 
George Loukas 

\thanks{S. Roy is with the University of Houston, Houston, TX, USA. E-mail: shantoroy@ieee.org. 
E. Panaousis, C. Noakes, S. Panda, and G. Loukas are with the Internet of Things and Security Centre, University of Greenwich, London SE10 9LS, UK. E-mail: \{e.panaousis, c.noakes, s.panda, g.loukas\}@greenwich.ac.uk.
A. Laszka is with the Pennsylvania State University, University Park, PA, USA. E-mail: aql5923@psu.edu.}%
\thanks{``This work has been submitted to IEEE S\&P 2024 for possible publication. Copyright may be transferred without notice, after which this version may no longer be accessible.''}
}

\pagestyle{plain}

\maketitle

\begin{abstract}
    The MITRE \attack framework, a comprehensive knowledge base of adversary tactics and techniques, has been widely adopted by the cybersecurity industry as well as by academic researchers. Its broad range of industry applications include threat intelligence, threat detection, and incident response, some of which go beyond what it was originally designed for. Despite its popularity, there is a lack of a systematic review of the applications and the research on \attack. This systematization of work aims to fill this gap. To this end, it introduces the first taxonomic systematization of the research literature on \attack, studies its degree of usefulness in different applications, and identifies important gaps and discrepancies in the literature to identify key directions for future work. The results of this work provide valuable insights for academics and practitioners alike, highlighting the need for more research on the practical implementation and evaluation of \attack. 
\end{abstract}

\frenchspacing

\pagenumbering{gobble} %

\section{Introduction} \label{sec:intro}
\noindent \attack presents a curated and actionable repository of adversarial Tactics, Techniques and Procedures (TTPs) \cite{georgiadou2021assessing} and details the characterization of adversary behavior after a successful system exploitation \cite{bodeau2018cyber}. The cybersecurity industry uses \attack for various applications including threat detection, adversary emulation, red teaming, behavioral analytics, defensive gap assessment, cyber threat intelligence (CTI) and threat modeling \cite{sans2022,choi2020expansion,al2020learning,pennington2019getting,alexander2020mitre}. 

At the same time, \attack is embraced in various domains, including ICS \cite{claroty2021} and Enterprise \cite{cylance2019}. Many vendors, including Cisco, Fortinet and Claroty have stated the importance of \attack in CTI and how security experts can align their research with \attack \cite{cisco2021,fortinet2021,claroty2021}. Cloud platforms such as Microsoft Azure Security have also been mapped to \attack using TTPs \cite{attackiq2021}. 
Even organizations like the North Atlantic Treaty Organization and the U.S. Department of Homeland Security have been using \attack for CTI and modeling~\cite{parmar2019use,fox2018enhanced}. 

There is a number of academic \cite{wagner2019cyber,schlette2021comparative,cascavilla2021cybercrime,ibrahim2020challenges,dutta2020overview,zibak2022threat,abu2018cyber,tounsi2018survey,mavroeidis2017cyber} and industrial \cite{brown20212021,brown2019evolution,shackleford2017cyber} surveys that present the state-of-the-art approaches in CTI and discuss the necessity and impact of \attack in CTI \cite{schlette2021comparative,tounsi2018survey,mavroeidis2017cyber}. There are also works that survey threat modeling approaches. For example, Tayouri et al.~\cite{tayouri2023survey} surveyed attack graph-based models that utilize or extend the MulVal method and mapped these MulVAL interaction rules to \attack techniques for evaluation in attack scenarios. Sadlek et al.~\cite{sadlek2022current} explored the current challenges of threat identification using public enumerations. The authors studied the usability of \attack for threat modeling. Bodeau et al.~\cite{bodeau2018cyber} discussed various security frameworks, including NIST $800-154$, STRIDE, DREAD, OCTAVE, TARA, TAL, STIX, CAPEC, alongside \attack for threat modeling and cybersecurity risk assessment purposes. These works are not systematizations highlighting a dearth of systematic research that addresses \attack use cases, application domains, and research methodologies. Our paper fills this gap by addressing the following research questions.

\textbf{RQ1}: \textit{How does the use of \attack contribute to cybersecurity research, and in what application domains and use cases has \attack been employed in the literature?}\newline
The aim of RQ1 is to determine the effectiveness of using \attack in creating novel and impactful research. Additionally, this inquiry may serve as a foundation for future studies exploring the application domains and use cases for which \attack has been investigated, thus expediting the learning curve and enhancing the framework's practicality. 
Our analysis reveals that \attack plays a critical role in cyber threat intelligence, intrusion detection and prevention, risk assessment and mitigation, red/purple team exercises and professional training. We also highlight the diverse application domains of \attack, which include enterprise networks, industrial control systems, IoT and mobile communication systems.

\textbf{RQ2.} \textit{How is \attack correlated, mapped, or integrated with other security frameworks in practice?}\newline
Understanding this correlation will illuminate the value of the framework for industrial applications, which frequently need to comply with various frameworks to meet cybersecurity requirements. This insight can clarify the possibility of integrating \attack and these frameworks into a unified global framework. 
Our investigation reveals that several studies have attempted to combine \attack with other security frameworks such as the cyber kill chain, NIST CSF, ISMS, CAPEC, D3FEND and Diamond models. The integration of these frameworks results in more comprehensive solutions enabling us, for example, to identify more effective sets of security controls.

\textbf{RQ3.}\textit{What are some examples of how industry utilizes \attack, and what research trends, as mentioned in \textbf{RQ1}, have not yet been observed in the industrial applications of \attack?}\newline
Understanding the gap between the use of \attack in industry and academia can motivate the adoption of research methods that are better suited to realistic environments, thereby enabling the development of solutions to emerging societal and industrial problems. 
Our findings show that academic researchers tend to use \attack to develop models for attack scenarios, analyze threat intelligence datasets and investigate system vulnerabilities using mathematical and statistical models. The utilization of \attack enables them to demonstrate the applicability of their work to real-world scenarios, providing a stronger basis for their proposals and facilitating the assessment of their research. In contrast, the industry focuses more on developing CTI tools and frameworks, evaluating products against \attack tactics and techniques, improving red or purple team exercises and providing offensive security training. 

\pagenumbering{arabic} %
\setcounter{page}{2} %

\textbf{RQ4.} \textit{What scientific methods have academic researchers employed to construct attack scenarios, models, or methods using \attack matrices?}\newline
By examining these methods, we enable researchers to identify areas where these scientific methods are not being utilized to their full potential and determine the shortage of research that employs equally appropriate methods. 
These scientific methods include machine learning (ML), natural language processing (as subfield of ML), probability theory, graph theory and game theory. We examine how the \attack framework has been implemented in different projects. Specifically, we analyze the testbed environments and tools that researchers have used to evaluate their work based on \attack and how they have applied these tools to achieve their objectives. Finally, we investigate the methods used to evaluate research that utilizes \attack. These methods include numerical or statistical, human-based and model-based evaluations.

The rest of this paper is structured as follows: Section \ref{sec:taxonomy} proposes a novel taxonomy of concepts used to answer the research questions. Section~\ref{sec:applications} outlines the application domains and use cases in which \attack has been used to address significant cybersecurity challenges and how \attack has been combined with other security frameworks (answering \textbf{RQ1-RQ3}). Section \ref{sec:res_approaches} describes the research approaches used in conjunction with \attack in the literature (answering \textbf{RQ4}). Finally, Section~\ref{sec:lessons} summarizes the key points of this systematization work, including the significance and limitations of \attack and offers suggestions for future work.

\section{Proposed Taxonomy}\label{sec:taxonomy}
\begin{figure*}[!t]
    \centering
    \includegraphics[width=0.9\textwidth]{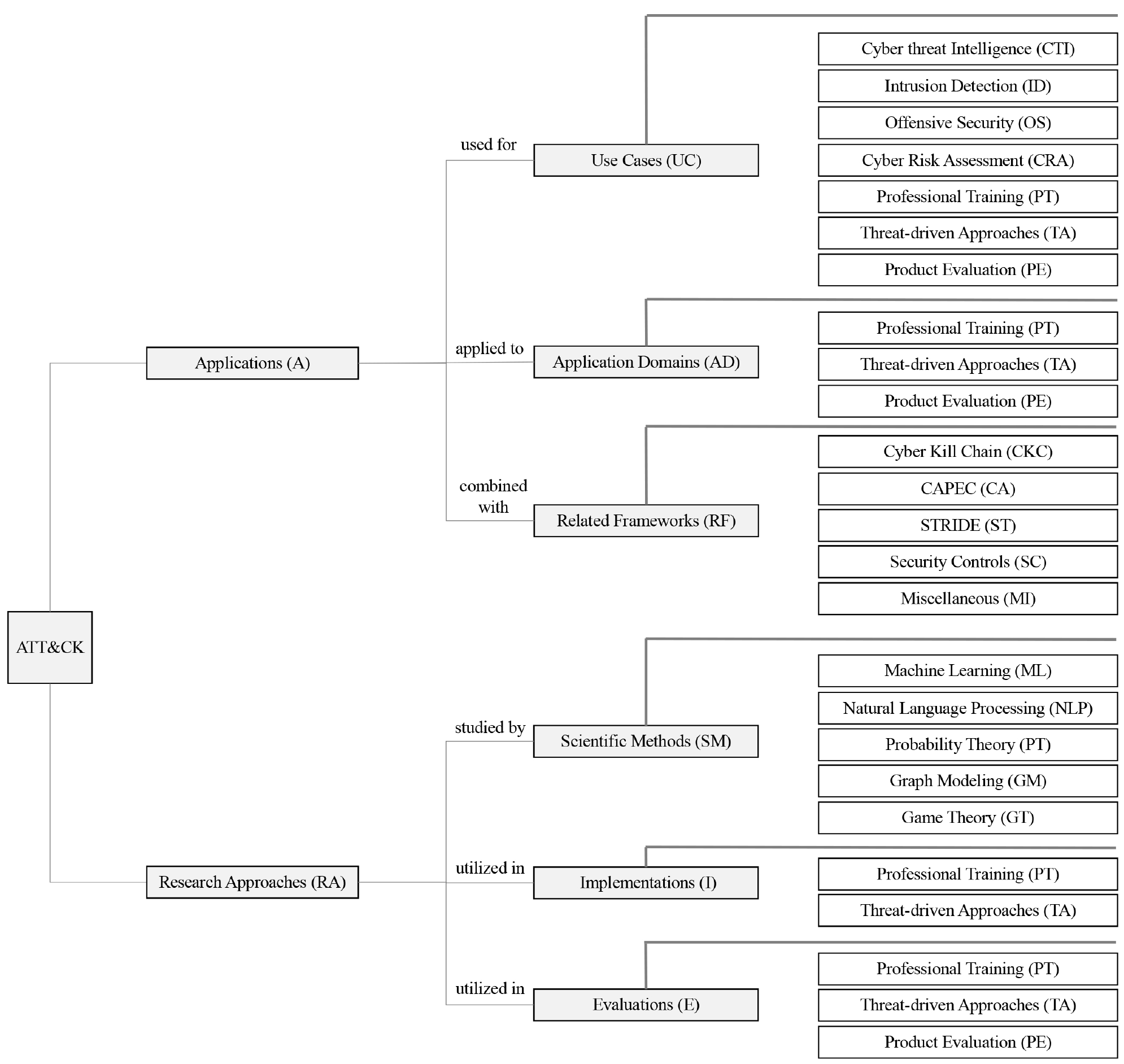}
    \caption{Taxonomy of papers that use \attack.}
    \label{fig:attack_survey}
\end{figure*}

\noindent To answer the research questions, we have defined a taxonomy (shown in Figure~\ref{fig:attack_survey}) that categorizes \attack-oriented applications, use cases and research approaches from the literature. By utilizing this taxonomy, researchers can classify the literature and gain insights into the usefulness of \attack and identify any gaps in research to date. Table~\ref{tab:use-of-attack} classifies all surveyed papers using our taxonomy. 

The first level classification in our taxonomy considers \attack-based Applications (A) and Research Approaches (RA) found in the literature. We divide \attack applications into three primary categories: Use Cases (UC), Application Domains (AD) and Related Frameworks (RF), with subcategories for each application. UC defines the specific applications where \attack has been utilized and is further categorized into Cyber Threat Intelligence (CTI), Intrusion Detection (ID), Offensive Security (OS), Cyber Risk Assessment (CRA), Professional Training (PT), Threat-driven Approaches (TA) and Product Evaluation (PE). AD defines the specific domains where \attack has been applied and we identify three application domains: Enterprise Networks (EN), Mobile Communication Systems (MCS) and Industrial Control Systems (ICS). We categorize RF into Cyber Kill Chain (CKC), CAPEC (CA), STRIDE (ST), Security Controls (SC) and the Miscellaneous (MI) subcategory for the rest of security frameworks.

We classify the research approaches in our study into three categories: Scientific Method (SM), Implementation (I) and Evaluation (E). SM identifies the research fields that used \attack in any capacity and we further categorize it into five subcategories: Machine Learning (ML), Natural Language Processing (NLP), Probability Theory (PT), Graph Modeling (GM) and Game Theory (GT). Implementation (I) defines how researchers utilized \attack in implementing the proposed works and we define two subcategories: Testbed (TE) and Tools (TO), developed to implement certain attack scenarios or models. Finally, we categorize Evaluation (E) into three subcategories: Numeric Evaluation (NE), Human Evaluation (HE) and Model Evaluation (ME). This category shows how researchers evaluated their testbeds, tools and models.

\begin{table*}[!ht]
\caption{Taxonomy classification of papers using \attack}
\label{tab:use-of-attack}
\resizebox{\textwidth}{!}{
\begin{tabular}{|c|ccc|ccc|}
\hline
\multirow{2}{*}{\diagbox[width=13.2em]{Literature}{Description}} & \multicolumn{3}{c|}{Application} & \multicolumn{3}{c|}{Research Approach} \\ \cline{2-7} 
 & \multicolumn{1}{c|}{Use-cases} & \multicolumn{1}{c|}{\begin{tabular}[c]{@{}c@{}}Application\\ Domain\end{tabular}} & \begin{tabular}[c]{@{}c@{}}Related\\ Frameworks\end{tabular} & \multicolumn{1}{c|}{\begin{tabular}[c]{@{}c@{}}Scientific\\ Methods\end{tabular}} & \multicolumn{1}{c|}{Implementation} & Evaluation \\ \hline\hline
 Ahn et al. (2020)~\cite{ahn2020research} & \multicolumn{1}{c|}{PT} & \multicolumn{1}{c|}{EN} & \xm & \multicolumn{1}{c|}{GM} & \multicolumn{1}{c|}{TE} & NE \\ \hline
 Ampel et al. (2021)~\cite{ampel2021linking} & \multicolumn{1}{c|}{OS} & \multicolumn{1}{c|}{EN} & \xm & \multicolumn{1}{c|}{ML} & \multicolumn{1}{c|}{TE} & NE,ME \\ \hline
 Choi et al. (2020)~\cite{choi2020expansion} & \multicolumn{1}{c|}{CTI} & \multicolumn{1}{c|}{ICS} & \xm & \multicolumn{1}{c|}{\xm} & \multicolumn{1}{c|}{TE, TO} & \xm \\ \hline
 Georgiadou et al. (2021)~\cite{georgiadou2021assessing} & \multicolumn{1}{c|}{PT, OS, CRA} & \multicolumn{1}{c|}{EN} & \xm & \multicolumn{1}{c|}{\xm} & \multicolumn{1}{c|}{\xm} & \xm \\ \hline
 Hong et al. (2019)~\cite{hong2019design}  & \multicolumn{1}{c|}{PT, OS} & \multicolumn{1}{c|}{EN} & \xm & \multicolumn{1}{c|}{ML} & \multicolumn{1}{c|}{TE} & \xm \\ \hline
 Kuppa et al. (2021)~\cite{kuppa2021linking} & \multicolumn{1}{c|}{OS} & \multicolumn{1}{c|}{EN} & \xm & \multicolumn{1}{c|}{NLP} & \multicolumn{1}{c|}{\xm} & ME, NE \\ \hline
 Munaiah et al. (2019)~\cite{munaiah2019characterizing} & \multicolumn{1}{c|}{OS} & \multicolumn{1}{c|}{EN} & \xm & \multicolumn{1}{c|}{GM} & \multicolumn{1}{c|}{\xm} & NE \\ \hline
 Outkin et al. (2021)~\cite{outkin2021defender}  & \multicolumn{1}{c|}{TA, PE} & \multicolumn{1}{c|}{EN} & \xm & \multicolumn{1}{c|}{GM} & \multicolumn{1}{c|}{\xm} & ME \\ \hline
 Pell et al. (2021)~\cite{pell2021towards} & \multicolumn{1}{c|}{TA} & \multicolumn{1}{c|}{MCS} & \xm & \multicolumn{1}{c|}{\xm} & \multicolumn{1}{c|}{\xm} & \xm \\ \hline
 Xiong et al. (2021)~\cite{xiong2021cyber}  & \multicolumn{1}{c|}{TA} & \multicolumn{1}{c|}{EN} & MI & \multicolumn{1}{c|}{GM} & \multicolumn{1}{c|}{TE} & ME \\ \hline 
 Hemberg et al. (2020)~\cite{hemberg2020linking} & \multicolumn{1}{c|}{CTI} & \multicolumn{1}{c|}{EN} & CA & \multicolumn{1}{c|}{GM} & \multicolumn{1}{c|}{TE} & NE \\ \hline
 Kim et al. (2021)~\cite{kim2021cyber} & \multicolumn{1}{c|}{PT, OS, PE} & \multicolumn{1}{c|}{EN} & CKC & \multicolumn{1}{c|}{GM} & \multicolumn{1}{c|}{\xm} & NE \\ \hline
 Choi et al. (2021)~\cite{choi2021probabilistic} & \multicolumn{1}{c|}{OS, PT} & \multicolumn{1}{c|}{ICS} & \xm & \multicolumn{1}{c|}{GM} & \multicolumn{1}{c|}{TO} & ME \\ \hline
 Al et al. (2020)~\cite{al2020learning} & \multicolumn{1}{c|}{CTI, OS} & \multicolumn{1}{c|}{EN} & \xm & \multicolumn{1}{c|}{ML,GM} & \multicolumn{1}{c|}{TO} & NE, HE \\ \hline
 Amro et al. (2021)~\cite{amro2021assessing} & \multicolumn{1}{c|}{CRA} & \multicolumn{1}{c|}{ICS} & CA & \multicolumn{1}{c|}{GM} & \multicolumn{1}{c|}{\xm} & NE \\ \hline
 Arshad et al. (2021)~\cite{arshad2021attack} & \multicolumn{1}{c|}{OS, PT} & \multicolumn{1}{c|}{EN} & \xm & \multicolumn{1}{c|}{GT} & \multicolumn{1}{c|}{TO} & ME \\ \hline
 Golushko et al. (2020)~\cite{golushko2020application} & \multicolumn{1}{c|}{IDS, OS} & \multicolumn{1}{c|}{EN, ICS, MCS} & \xm & \multicolumn{1}{c|}{\xm} & \multicolumn{1}{c|}{TO} & \xm \\ \hline
 Kriaa et al. (2021)~\cite{kriaa2021seckg} & \multicolumn{1}{c|}{CTI} & \multicolumn{1}{c|}{EN, ICS, MCS} & CA & \multicolumn{1}{c|}{GM} & \multicolumn{1}{c|}{TO} & ME \\ \hline
 Kwon et al. (2020)~\cite{kwon2020cyber} & \multicolumn{1}{c|}{CTI, TA} & \multicolumn{1}{c|}{ICS} & MI, CKC & \multicolumn{1}{c|}{\xm} & \multicolumn{1}{c|}{TO} & \xm \\ \hline
 Bromander et al. (2020)~\cite{bromander2020modeling} & \multicolumn{1}{c|}{CTI} & \multicolumn{1}{c|}{ICS} & CA & \multicolumn{1}{c|}{GM} & \multicolumn{1}{c|}{TO} & ME \\ \hline
 Legoy et al. (2020)~\cite{legoy2020automated} & \multicolumn{1}{c|}{CTI} & \multicolumn{1}{c|}{EN, MCS, ICS} & \xm & \multicolumn{1}{c|}{ML} & \multicolumn{1}{c|}{TO} & ME \\ \hline
 Fairbanks et al. (2021)~\cite{fairbanks2021att} & \multicolumn{1}{c|}{CTI} & \multicolumn{1}{c|}{MCS} & \xm & \multicolumn{1}{c|}{GM, ML} & \multicolumn{1}{c|}{\xm} & NE \\ \hline
 Huang et al. (2021)~\cite{huang2021open} & \multicolumn{1}{c|}{CTI, TA} & \multicolumn{1}{c|}{EN, MCS, ICS} & MI & \multicolumn{1}{c|}{ML} & \multicolumn{1}{c|}{TO} & NE, ME \\ \hline
 Kurniawan et al. (2021)~\cite{kurniawan2021att} & \multicolumn{1}{c|}{CTI} & \multicolumn{1}{c|}{EN, MCS, ICS} & MI & \multicolumn{1}{c|}{GM} & \multicolumn{1}{c|}{TO} & ME \\ \hline
 Lakhdhar et al. (2021)~\cite{lakhdhar2021machine} & \multicolumn{1}{c|}{CTI, TA} & \multicolumn{1}{c|}{EN, MCS, ICS} & CA, MI & \multicolumn{1}{c|}{ML} & \multicolumn{1}{c|}{TO} & NE, ME \\ \hline
 Lee et al. (2021)~\cite{lee2021fileless} & \multicolumn{1}{c|}{CTI} & \multicolumn{1}{c|}{EN, MCS, ICS} & \xm & \multicolumn{1}{c|}{GM, ML} & \multicolumn{1}{c|}{TO} & NE, ME \\ \hline
 Mendsaikhan et al. (2020)~\cite{mendsaikhan2020automatic} & \multicolumn{1}{c|}{CTI, TA} & \multicolumn{1}{c|}{EN, MCS, ICS} & CA & \multicolumn{1}{c|}{ML} & \multicolumn{1}{c|}{TO} & NE, ME \\ \hline
 Parmar et al. (2019)~\cite{parmar2019use} & \multicolumn{1}{c|}{CTI} & \multicolumn{1}{c|}{EN, MCS, ICS} & \xm & \multicolumn{1}{c|}{GM} & \multicolumn{1}{c|}{TE, TO} & NE \\ \hline
 Purba et al. (2020)~\cite{purba2020word} & \multicolumn{1}{c|}{CTI} & \multicolumn{1}{c|}{EN, MCS, ICS} & \xm & \multicolumn{1}{c|}{NLP} & \multicolumn{1}{c|}{TO} & NE, ME \\ \hline
 Aghaei et al. (2019)~\cite{aghaei2019threatzoom} & \multicolumn{1}{c|}{CTI, TA} & \multicolumn{1}{c|}{EN, MCS, ICS} & CA, MI & \multicolumn{1}{c|}{\xm} & \multicolumn{1}{c|}{\xm} & \xm \\ \hline
 Ajmal et al. (2021)~\cite{ajmal2021offensive} & \multicolumn{1}{c|}{CTI, OS, PT} & \multicolumn{1}{c|}{EN} & \xm & \multicolumn{1}{c|}{PT} & \multicolumn{1}{c|}{TE, TO} & NE, ME \\ \hline
 Brazhuk et al. (2021)~\cite{brazhuk2021towards} & \multicolumn{1}{c|}{CTI, TA} & \multicolumn{1}{c|}{EN, MCS, ICS} & CA, ST, SC, MI & \multicolumn{1}{c|}{GM, PT} & \multicolumn{1}{c|}{TO} & NE \\ \hline
 Elitzur et al. (2019)~\cite{elitzur2019attack} & \multicolumn{1}{c|}{CTI, TA} & \multicolumn{1}{c|}{EN, MCS, ICS} & SC & \multicolumn{1}{c|}{GM, PT} & \multicolumn{1}{c|}{TO} & NE, ME \\ \hline
 Fairbanks et al. (2021)~\cite{fairbanks2021identifying} & \multicolumn{1}{c|}{CTI} & \multicolumn{1}{c|}{EN, MCS, ICS} & \xm & \multicolumn{1}{c|}{GM, ML} & \multicolumn{1}{c|}{\xm} & NE, ME \\ \hline
 Franklin et al. (2017)~\cite{franklin2017toward} & \multicolumn{1}{c|}{TA} & \multicolumn{1}{c|}{EN, MCS, ICS} & \xm & \multicolumn{1}{c|}{ML} & \multicolumn{1}{c|}{TO} & \xm \\ \hline
 Gourisetti et al. (2019)~\cite{gourisetti2019demonstration} & \multicolumn{1}{c|}{CTI, TA, CRA} & \multicolumn{1}{c|}{ICS} & SC, MI & \multicolumn{1}{c|}{\xm} & \multicolumn{1}{c|}{TE, TO} & NE \\ \hline
 Gylling et al. (2021)~\cite{gylling2021mapping} & \multicolumn{1}{c|}{CTI, TA} & \multicolumn{1}{c|}{EN, MCS, ICS} & \xm & \multicolumn{1}{c|}{GM} & \multicolumn{1}{c|}{TO} & NE \\ \hline
 Hacks et al. (2021,2022)~\cite{hacks2021integrating, hacks2022measuring} & \multicolumn{1}{c|}{CTI, TA} & \multicolumn{1}{c|}{EN, MCS, ICS} & SC & \multicolumn{1}{c|}{\xm} & \multicolumn{1}{c|}{TO} & NE, ME, HE \\ \hline
 Hassanzadeh et al. (2020)~\cite{hassanzadeh2018samiit} & \multicolumn{1}{c|}{CTI, TA} & \multicolumn{1}{c|}{ICS} & SC & \multicolumn{1}{c|}{ML} & \multicolumn{1}{c|}{TE, TO} & NE, ME \\ \hline
 Ahmed et al. (2022)~\cite{ahmed2022mitre} & \multicolumn{1}{c|}{CRA, TA} & \multicolumn{1}{c|}{EN, MCS, ICS} & SC & \multicolumn{1}{c|}{PT, GM} & \multicolumn{1}{c|}{\xm} & ME \\ \hline
 Bolbot et al. (2022)~\cite{bolbot2022investigating} & \multicolumn{1}{c|}{CRA, TA} & \multicolumn{1}{c|}{EN, MCS, ICS} & ST, MI & \multicolumn{1}{c|}{\xm} & \multicolumn{1}{c|}{TO} & NE \\ \hline
 Oruc et al. (2022)~\cite{oruc2022assessing} & \multicolumn{1}{c|}{CRA} & \multicolumn{1}{c|}{ICS} & SC & \multicolumn{1}{c|}{PT} & \multicolumn{1}{c|}{TO} & NE \\ \hline
 TJ OConnor (2022)~\cite{oconnor2022helo} & \multicolumn{1}{c|}{PT} & \multicolumn{1}{c|}{EN} & \xm & \multicolumn{1}{c|}{\xm} & \multicolumn{1}{c|}{TE, TO} & NE, HE \\ \hline
Kim et al. (2020)~\cite{kim2020automated} & \multicolumn{1}{c|}{PT, PE} & \multicolumn{1}{c|}{EN} & CKC & \multicolumn{1}{c|}{GM} & \multicolumn{1}{c|}{TE, TO} & ME \\ \hline
Sadlek et al. (2022)~\cite{sadlek2022current} & \multicolumn{1}{c|}{CTI, TA} & \multicolumn{1}{c|}{EN} & CA, MI & \multicolumn{1}{c|}{\xm} & \multicolumn{1}{c|}{\xm} & NE \\ \hline
Rao et al. (2023)~\cite{rao2023threat} & \multicolumn{1}{c|}{TA} & \multicolumn{1}{c|}{MCS} & MI & \multicolumn{1}{c|}{GM} & \multicolumn{1}{c|}{\xm} & ME \\ \hline
Chen et al. (2022)~\cite{chen2022building} & \multicolumn{1}{c|}{TA, PE} & \multicolumn{1}{c|}{EN} & MI & \multicolumn{1}{c|}{NLP, GM} & \multicolumn{1}{c|}{TO} & NE \\ \hline
Adam et al. (2022)~\cite{adam2022attack} & \multicolumn{1}{c|}{CTI} & \multicolumn{1}{c|}{EN} & CA, MI & \multicolumn{1}{c|}{NLP, ML} & \multicolumn{1}{c|}{\xm} & NE, ME \\ \hline
Sadlek et al. (2022)~\cite{sadlek2022identification} & \multicolumn{1}{c|}{CTI, TA} & \multicolumn{1}{c|}{EN, MCS, ICS} & ST, CKC, SC & \multicolumn{1}{c|}{GM} & \multicolumn{1}{c|}{TE, TO} & ME \\ \hline
Jadidi et al. (2021)~\cite{jadidi2021threat} & \multicolumn{1}{c|}{CTI, TA} & \multicolumn{1}{c|}{ICS} & SC & \multicolumn{1}{c|}{GM} & \multicolumn{1}{c|}{TO} & ME \\ \hline
Mundt et al. (2022)~\cite{mundt2022towards} & \multicolumn{1}{c|}{CTI} & \multicolumn{1}{c|}{EN} & MI & \multicolumn{1}{c|}{GM} & \multicolumn{1}{c|}{TO} & ME \\ \hline
Niakanlahiji et al. (2018)~\cite{niakanlahiji2018natural} & \multicolumn{1}{c|}{CTI} & \multicolumn{1}{c|}{EN} & MI & \multicolumn{1}{c|}{NLP} & \multicolumn{1}{c|}{TO} & NE, ME \\ \hline
Ayoade et al. (2018)~\cite{ayoade2018automated} & \multicolumn{1}{c|}{CTI} & \multicolumn{1}{c|}{EN} & CKC, MI & \multicolumn{1}{c|}{NLP, PT, ML} & \multicolumn{1}{c|}{TO} & NE, ME \\ \hline
Karuna et al. (2021)~\cite{karuna2021automating} & \multicolumn{1}{c|}{CTI} & \multicolumn{1}{c|}{EN, MCS, ICS} & \xm & \multicolumn{1}{c|}{NLP} & \multicolumn{1}{c|}{\xm} & \xm \\ \hline
Shin et al. (2021)~\cite{shin2021art} & \multicolumn{1}{c|}{CTI} & \multicolumn{1}{c|}{EN, MCS, ICS} & \xm & \multicolumn{1}{c|}{PT, ML} & \multicolumn{1}{c|}{TO} & NE \\ \hline
He et al. (2021)~\cite{he2021model} & \multicolumn{1}{c|}{CTI, CRA} & \multicolumn{1}{c|}{EN, MCS, ICS} & MI & \multicolumn{1}{c|}{PT, GM} & \multicolumn{1}{c|}{\xm} & NE \\ \hline
Johnson et al. (2018)~\cite{johnson2018meta} & \multicolumn{1}{c|}{TA, OS} & \multicolumn{1}{c|}{EN, MCS, ICS} & SC, MI & \multicolumn{1}{c|}{PT, GM} & \multicolumn{1}{c|}{TO} & NE \\ \hline
Tayouri et al. (2023)~\cite{tayouri2023survey} & \multicolumn{1}{c|}{TA, CTI} & \multicolumn{1}{c|}{EN} & MI & \multicolumn{1}{c|}{GM} & \multicolumn{1}{c|}{TO} & NE \\ \hline
Bodeau et al. (2018)~\cite{bodeau2018cyber} & \multicolumn{1}{c|}{CTI, TA, CRA} & \multicolumn{1}{c|}{EN, MCS, ICS} & CA, CKC, ST, SC, MI & \multicolumn{1}{c|}{\xm} & \multicolumn{1}{c|}{\xm} & \xm \\ \hline
Manocha et al. (2021)~\cite{manocha2021security} & \multicolumn{1}{c|}{CRA, OS} & \multicolumn{1}{c|}{EN, MCS, ICS} & MI & \multicolumn{1}{c|}{PT} & \multicolumn{1}{c|}{\xm} & NE \\ \hline
Mashima et al. (2022)~\cite{mashima2022mitre} & \multicolumn{1}{c|}{CTI, PE} & \multicolumn{1}{c|}{ICS} & SC, MI & \multicolumn{1}{c|}{\xm} & \multicolumn{1}{c|}{TE, TO} & ME \\ \hline
Dhirani et al. (2021)~\cite{dhirani2021industrial} & \multicolumn{1}{c|}{TA, PE, CRA} & \multicolumn{1}{c|}{ICS} & SC, MI & \multicolumn{1}{c|}{\xm} & \multicolumn{1}{c|}{\xm} & \xm \\ \hline
Luh et al. (2022)~\cite{luh2022penquest} & \multicolumn{1}{c|}{TA, OS} & \multicolumn{1}{c|}{EN, ICS} & CA, SC, MI & \multicolumn{1}{c|}{GT} & \multicolumn{1}{c|}{TE, TO} & HE, NE \\ \hline
Husari et al. (2019)~\cite{husari2019learning} & \multicolumn{1}{c|}{CTI} & \multicolumn{1}{c|}{EN, MCS, ICS} & CKC, MI & \multicolumn{1}{c|}{NLP} & \multicolumn{1}{c|}{TO} & \xm \\ \hline
Nisioti et al. (2021)~\cite{nisioti2021game} & \multicolumn{1}{c|}{TA, OS} & \multicolumn{1}{c|}{EN, MCS, ICS} & CKC, MI & \multicolumn{1}{c|}{GT, GM} & \multicolumn{1}{c|}{TO} & NE, ME \\ \hline
Halvorsen et al. (2019)~\cite{halvorsen2019evaluating} & \multicolumn{1}{c|}{CTI, TA, ID} & \multicolumn{1}{c|}{EN} & SC, MI & \multicolumn{1}{c|}{PT} & \multicolumn{1}{c|}{TE, TO} & NE \\ \hline
\end{tabular}
}
\end{table*}

\begin{table*}[!ht]
\ContinuedFloat 
\caption{Taxonomy classification of papers using \attack (continued)}
\label{tab:use-of-attack}
\resizebox{\textwidth}{!}{
\begin{tabular}{|c|ccc|ccc|}
\hline
\multirow{2}{*}{\diagbox[width=12em]{Literature}{Description}} & \multicolumn{3}{c|}{Application} & \multicolumn{3}{c|}{Research Approach} \\ \cline{2-7} 
 & \multicolumn{1}{c|}{Use-cases} & \multicolumn{1}{c|}{\begin{tabular}[c]{@{}c@{}}Application\\ Domain\end{tabular}} & \begin{tabular}[c]{@{}c@{}}Related\\ Frameworks\end{tabular} & \multicolumn{1}{c|}{\begin{tabular}[c]{@{}c@{}}Scientific\\ Methods\end{tabular}} & \multicolumn{1}{c|}{Implementation} & Evaluation \\ \hline\hline
Wong et al. (2021)~\cite{wong2021threat} & \multicolumn{1}{c|}{TA, OS} & \multicolumn{1}{c|}{EN} & ST & \multicolumn{1}{c|}{\xm} & \multicolumn{1}{c|}{\xm} & \xm \\ \hline
Dhir et al. (2021)~\cite{dhir2021prospective} & \multicolumn{1}{c|}{TA, OS} & \multicolumn{1}{c|}{EN} & \xm & \multicolumn{1}{c|}{PT} & \multicolumn{1}{c|}{\xm} & ME \\ \hline
Holder et al. (2021)~\cite{holder2021explainable} & \multicolumn{1}{c|}{TA, CRA} & \multicolumn{1}{c|}{EN} & \xm & \multicolumn{1}{c|}{PT} & \multicolumn{1}{c|}{TE} & NE \\ \hline
Ahn et al. (2022)~\cite{ahn2022malicious} & \multicolumn{1}{c|}{ID, CTI, TA} & \multicolumn{1}{c|}{EN} & \xm & \multicolumn{1}{c|}{PT, GM, NLP} & \multicolumn{1}{c|}{TE, TO} & NE \\ \hline
Stoleriu et al. (2021)~\cite{stoleriu2021cyber} & \multicolumn{1}{c|}{ID} & \multicolumn{1}{c|}{EN} & SC, MI & \multicolumn{1}{c|}{ML} & \multicolumn{1}{c|}{TE, TO} & NE \\ \hline
Bagui et al. (2022)~\cite{bagui2022detecting} & \multicolumn{1}{c|}{ID} & \multicolumn{1}{c|}{EN} & \xm & \multicolumn{1}{c|}{ML} & \multicolumn{1}{c|}{\xm} & NE, ME \\ \hline
Zurowski et al. (2018)~\cite{zurowski2022quantitative} & \multicolumn{1}{c|}{TA, OS} & \multicolumn{1}{c|}{EN} & SC & \multicolumn{1}{c|}{ML} & \multicolumn{1}{c|}{TO} & NE \\ \hline
Alnafrani et al. (2022)~\cite{alnafrani2022aifis} & \multicolumn{1}{c|}{TA} & \multicolumn{1}{c|}{EN} & \xm & \multicolumn{1}{c|}{PT, ML} & \multicolumn{1}{c|}{TE, TO} & ME \\ \hline
Samtani et al. (2022)~\cite{samtani2022explainable} & \multicolumn{1}{c|}{CTI, TA} & \multicolumn{1}{c|}{EN} & \xm & \multicolumn{1}{c|}{ML} & \multicolumn{1}{c|}{\xm} & \xm \\ \hline
Grigorescu et al. (2022)~\cite{grigorescu2022cve2att} & \multicolumn{1}{c|}{CTI} & \multicolumn{1}{c|}{EN} & CA, MI & \multicolumn{1}{c|}{NLP, ML, GM} & \multicolumn{1}{c|}{TO} & NE, ME \\ \hline
Hasan et al. (2019)~\cite{hasan2019artificial} & \multicolumn{1}{c|}{ID} & \multicolumn{1}{c|}{EN, ICS} & CA, CKC, MI & \multicolumn{1}{c|}{ML, GM} & \multicolumn{1}{c|}{TO} & NE, ME \\ \hline
Maymí et al. (2017)~\cite{maymi2017towards} & \multicolumn{1}{c|}{CTI} & \multicolumn{1}{c|}{EN, MCS, ICS} & CKC, MI & \multicolumn{1}{c|}{ML, GM} & \multicolumn{1}{c|}{\xm} & \xm \\ \hline
Drašar et al. (2020)~\cite{dravsar2020session} & \multicolumn{1}{c|}{TA, OS} & \multicolumn{1}{c|}{EN, MCS, ICS} & MI & \multicolumn{1}{c|}{GT, GM} & \multicolumn{1}{c|}{TE, TO} & NE, ME \\ \hline
Kim et al. (2022)~\cite{kim2022comparative} & \multicolumn{1}{c|}{CTI, TA} & \multicolumn{1}{c|}{EN} & ST, CKC & \multicolumn{1}{c|}{ML} & \multicolumn{1}{c|}{TO} & NE, ME \\ \hline
Kim et al. (2021)~\cite{kim2021automatically} & \multicolumn{1}{c|}{CTI, TA} & \multicolumn{1}{c|}{MCS} & \xm & \multicolumn{1}{c|}{ML} & \multicolumn{1}{c|}{\xm} & ME, NE \\ \hline
Sahu et al. (2021)~\cite{sahu2021model} & \multicolumn{1}{c|}{TA} & \multicolumn{1}{c|}{EN} & MI & \multicolumn{1}{c|}{\xm} & \multicolumn{1}{c|}{\xm} & ME \\ \hline
Zhao et al. (2021,2022)~\cite{zhao2021exploring,zhao2022cloud} & \multicolumn{1}{c|}{PT} & \multicolumn{1}{c|}{EN} & MI & \multicolumn{1}{c|}{\xm} & \multicolumn{1}{c|}{\xm} & \xm \\ \hline
Van et al. (2022)~\cite{van2022identifying} & \multicolumn{1}{c|}{TA} & \multicolumn{1}{c|}{EN} & MI, SC & \multicolumn{1}{c|}{PT, GM} & \multicolumn{1}{c|}{TO} & ME \\ \hline
\end{tabular}
}
\end{table*}

\section{Applications (A) of \attack} \label{sec:applications}
\noindent In this section, we explore \attack's application domains and use cases, examine how other security frameworks are mapped to or combined with \attack and finally delve into the different ways in which \attack is utilized by academia and industry. 

\vspace{-0.2cm}
\subsection{A-UC: Use Cases}

\subsubsection{A-UC-CTI: Cyber Threat Intelligence}

According to Legoy et al. \cite{legoy2020automated}, Cyber Threat Intelligence (CTI) is a continuous process that necessitates the use of text classification techniques for retrieving TTP-oriented information. 
Mundt et al. \cite{mundt2022towards} combined CTI with Information Security Management Systems (ISMS) and automate CTI by utilizing \attack. Al et al.~\cite{al2020learning} examined the connections between \attack techniques enabling the prediction of previously unobserved ones. Kriaa et al.~\cite{kriaa2021seckg} used \attack to create their detection and prediction module by constructing a knowledge graph of TTPs. Zhang et al.~\cite{zhang2022automatic} proposed a model that uses \attack to assess CTI reports automatically to extract Indicators of Compromise (IoC) timely. Here, the authors used \attack to identify attack techniques related to IoC.

The increasing number of connected IoT devices, which bear security vulnerabilities, is contributing to the constantly evolving operational technology (OT) cyber threat landscape. To address this issue, Kwon et al. \cite{kwon2020cyber} developed a Cyber Threat Dictionary utilizing the \attack ICS matrix and mapped security controls to the \attack ICS matrix. Odemis et al.~\cite{odemis2022detecting} utilized \attack to create a cyber expertise test to detect and categorize adversarial behavior for their CTI research. Similarly, \attack was also used for further threat analysis and adversarial TTP classification in the works of Lee et al.~\cite{lee2021fileless}, Mendsaikhan et al.~\cite{mendsaikhan2020automatic} and Jo et al.~\cite{jo2022cyberattack}. Hemberg et al.~\cite{hemberg2020linking} and Kurniawan et al.~\cite{kurniawan2021att} utilized the framework for linking \attack techniques to vulnerabilities. Additionally, Bromander et al.~\cite{bromander2020modeling} developed a CTI data model that identifies threats with \attack being used as a source for tactics, techniques, tools, and threat actors.

\textbf{Insight 1.} \attack is valuable for security teams seeking to keep up with the latest threats and enhance their CTI capabilities. Many studies link CTI reports with \attack matrices to create effective mitigation strategies. It has been observed that the majority of research papers in this field utilize \attack to improve CTI. However, there is a lack of investigation into how insights gained from CTI research can be used to enhance \attack itself.

\subsubsection{A-UC-ID: Intrusion Detection}
\attack techniques can be used to categorize adversary behavior and detect advanced intrusions \cite{kuppa2021linking}. Common Vulnerabilities and Exposures (CVEs) can be linked to specific exploitation strategies and then mapped to \attack techniques. 

Golushko et al.~\cite{golushko2020application} applied \attack to identify effective techniques under the \textit{Command and Control} and \textit{Defense Evasion} tactics and provided recommendations for detection and prevention. Kriaa et al.~\cite{kriaa2021seckg} proposed a novel approach for building knowledge graphs using \attack and utilizing prediction techniques on event logs to identify and prevent 5G radio access network attacks. Additionally, Kwon et al.~\cite{kwon2020cyber} extended the \attack ICS matrix leading to the creation of new categories for threat detection and mitigation. 

The \dettect (Detect Tactics, Techniques \& Combat Threats) framework \cite{dettectgithub,DeTTCTMa95:online} was introduced by the industry \cite{roberts2021structured} to enhance intrusion detection. \dettect helps blue teams evaluate and analyze the quality and visibility of data log sources and detection coverage using \attack. 
In addition to \dettect, the industry is continuously creating frameworks and tools for detecting and responding to security incidents. These frameworks and tools are exemplified by \attack. For instance, Security Information and Event Management (SIEM) tools are adopting \attack for better detection and alert management \cite{hassanzadeh2018samiit,farooq2018optimal,HowMITRE75:online,pennington2019getting}.

\textbf{Insight 2.} \attack assists researchers identify behavior patterns of known threats and recognize the use of particular techniques and tools, which can aid in intrusion detection. There is a lack of studies that evaluate the effectiveness of \attack in supporting intrusion detection frameworks in real-world settings as well as research on how to adapt \attack to detect new threats.

\subsubsection{A-UC-OS: Offensive Security}
\attack is a valuable resource to conduct effective adversary emulation as it contains comprehensive information on techniques employed by various threat actors. By utilizing the \attack knowledge base, organizations can simulate realistic attack scenarios and proactively identify potential security gaps thus enhancing their overall security posture.

Kuppa et al.~\cite{kuppa2021linking} leveraged a CVE regular expression dataset to identify frequently exploited CVEs created by collecting different APT reports\footnote{\url{https://github.com/CyberMonitor/APT_CyberCriminal_Campagin_Collections}} from 2008 to 2019, zero-day exploits\footnote{\url{https://googleprojectzero.blogspot.com/p/0day.html}} from Google project zero, 63720 vulnerability reports and 37000 threat reports\footnote{\url{https://www.broadcom.com/support/security-center/a-z}}. 
The researchers obtained a sample of 200 CVEs from publicly available threat reports along with their corresponding \attack techniques to extract the relevant context phrases. Munaiah et al.~\cite{munaiah2019characterizing} used data from the 2018 National Collegiate Penetration Testing Competition and codified their approach in terms of \attack tactics and techniques that it is possible to characterize attacker campaigns as a chronological series of them.

Kim et al. \cite{kim2021cyber} developed an offensive security taxonomy and provided a systematic cyber attack scoring model. They employed artifacts from attacks to identify the techniques used. They constructed the technology and stages used by malware based on \attack and grouped the identified attack techniques used in a few real cyber-attack incidents. Other studies such as \cite{ampel2021linking,al2020learning,arshad2021attack,ajmal2021offensive} have also utilized \attack for offensive security practices, red teaming exercises, and penetration testing.

\textbf{Insight 3.} \attack is a valuable tool for offensive security teams to plan and execute simulated attacks in order to test an organization's security measures. By utilizing \attack, organizations can identify weaknesses in their defenses and improve their overall security posture. This proactive security approach can help organizations better protect themselves from real-world attacks. A potential research gap in the application of \attack to offensive security is the development of metrics for evaluating the effectiveness of defensive measures against specific TTPs, as well as standardized methods for mapping defensive measures to specific TTPs.

\subsubsection{A-UC-CRA: Cyber Risk Assessment}
ISO 27005, COBIT 5, NIST SP 800-30 and other frameworks are widely used for cyber risk assessment. Researchers have recently combined these frameworks with \attack for more effective risk assessment. For example, Ahmed et al.~\cite{ahmed2022mitre} proposed a methodology that uses \attack, NIST SP 800-30 Rev.1, and attack graphs to assess and characterize cyber risk.  Sadrazamis \cite{sadrazamis2022mitre} proposed a hierarchical risk assessment system based on \attack knowledge graph. Amro et al. \cite{amro2023assessing} employed semantics and components of \attack to quantify risks associated with cyber-physical systems. Ahmed et al. \cite{ahmed2022mitre} analyzed and characterized TTPs used by different threat actors for informed cyber risk assessment. In their study, Kure et al. \cite{kure2022integrated} presented an integrated cyber risk management framework that utilizes an \attack-driven threat modeling approach. Oruc et al. \cite{oruc2022assessing} used \attack to assess risks associated with cyber threats and vulnerabilities for integrated navigation systems on board shipping vessels. 

The use of cyber risk and vulnerability assessment data mapped to \attack tactics and techniques has been explored by the industry to identify mitigation strategies \cite{RiskandV25:online,RVAsMapp12:online}. Grantek \cite{grantek12:online} has detailed an approach to utilizing ICS \attack strategies for risk management, which involves system identification and characterization, vulnerability identification and threat modeling, and risk calculation and management. AttackIQ, a security research organization focused on prioritizing vulnerability management, published a whitepaper proposing the use of \attack and CVE for better risk management \cite{ciso12:online}. MITRE presented cyber resiliency metrics and scoring for better risk management in a whitepaper by Bodeau et al. \cite{bodeau2018cyber}. 

On the research front, Georgiadou et al. \cite{georgiadou2021assessing} associated individual and organizational culture dimensions with adversarial behavior and patterns documented in \attack, using a cybersecurity culture framework. They developed a hybrid \attack for Enterprise and ICS matrix to identify cyber risks to which an organization lacks resilience. 

\textbf{Insight 4.} By mapping threat behaviors with vulnerabilities, researchers have been able to provide essential mitigation tactics for assessing cyber risk. \attack provides a consistent and repeatable approach to evaluating security risks, enabling organizations to make more informed decisions about the threats they are facing. Investigating further the integration of \attack with other risk assessment methodologies has the potential to enhance its utility and effectiveness. Additionally, conducting empirical studies in collaboration with practitioners to assess the impact of using \attack on cyber risk quantification can contribute to advancing cyber risk management practice.

\subsubsection{A-UC-PT: Professional Training}
Georgiadou et al.~\cite{georgiadou2021assessing} focused on cyber warfare simulations for training offense and defense from real-world cyber scenarios related to \attack. Hong et al.~\cite{hong2019design} proposed an automated script to generate simulated threats for training professionals with practical methods for real-world defensive scenarios. O'Connor~\cite{oconnor2022helo} shared experiences, lessons and materials from an undergraduate course that suggests using \attack to combine theoretical learning and exploratory labs. 

Kim et al.~\cite{kim2020automated} analyzed real-world data from \attack to propose CyTEA, a model that can generate simulated cyber threats for a cybersecurity training system. The simulation level was evaluated based on procedural, environmental and consequential similarities to determine if the model is suitable for real-world use and acceptable for industry usage. Arshad et al. \cite{arshad2021attack} proposed an attack specific language (ASL) based on \attack that is used to streamline and automate the functions of a cyber range, which is used for training. The authors used \attack to specify procedure classification and map corresponding tactics and techniques. Other researchers have also utilized \attack to improve or design professional training programs, including Ahn et al.~\cite{ahn2020research} and Ajmal et al.~\cite{ajmal2021offensive}.

\textbf{Insight 5.} \attack enables professionals to enhance their knowledge of the threat landscape and improve their hands-on skills in responding to cyber attacks. By staying up-to-date with the latest threats and using \attack to develop effective defensive strategies, security professionals can better protect their organizations against cyber threats. While there are studies proposing different ways to use \attack in training, a research gap exists in the form of comprehensive evaluation studies that assess the effectiveness and efficiency of these programs. Therefore, there is a need for more empirical research to compare and measure the impact of different approaches on the development of cybersecurity skills and knowledge.

\subsubsection{A-UC-TA: Threat-driven Approaches}
Jadidi et al.~\cite{jadidi2021threat} emphasized that threat hunting and modeling rely on various inputs, such as CTI, third-party notifications and data from security analysts, to identify threat actor behavior or vulnerabilities. In this way, security professionals can be empowered to stay ahead of emerging threats. To enhance mitigation strategies, Ampel et al. \cite{ampel2021linking} developed a model that automates the mapping of CVEs to \attack techniques within the matrix. They extracted data from 24,863 CVEs across various exploitation databases. Hacks et al. \cite{hacks2021integrating} proposed a solution that offers CTI capabilities by utilizing \attack and mapping its components to attack graphs labeled with CTI. Rao et al.~\cite{rao2023threat} introduced a threat modeling framework named \emph{Bhadra}, designed specifically for MCS. Bhadra aligns with \attack for enterprise networks and can be used with or without \attack for threat modeling purposes. Since there was not any dedicated threat modeling framework for MCS, the authors looked into \attack for Enterprise and reused the structure and terminology of \attack.

Sadleck et al.~\cite{sadlek2022current} introduced an approach for managing and modeling threats by leveraging Common Platform Enumeration for asset management, CVEs and CWE for vulnerability management, and CAPEC and \attack for threat management. By using \attack and CAPEC together, the authors were able to provide a comprehensive description of adversarial tactics and techniques and attack patterns, leading to better threat management. Kim et al.~\cite{kim2021automatically} proposed an automated framework for attributing mobile threat actors by analyzing the mobile malware using automated \attack-based TTP and Indicators of Compromise. Similarly, Fox et al.~\cite{fox2018enhanced} developed an enhanced cyber threat model for the financial service sector that utilizes \attack and CAPEC. In another study, Jadidi et al.~\cite{jadidi2021threat} presented a threat-hunting framework to detect cyber threats against ICS devices during the early stages of the attack lifecycle. The authors leveraged \attack to generate hunting hypotheses and predict the future behavior of potential adversaries.

Numerous other research papers have used \attack for threat modeling \cite{chen2022building,kwon2020cyber,huang2021open,lakhdhar2021machine,mendsaikhan2020automatic,aghaei2019threatzoom,brazhuk2021towards,elitzur2019attack,gourisetti2019demonstration,gylling2021mapping,hacks2021integrating,hassanzadeh2018samiit,xiong2022cyber}. Among these, Elitzur et al.~\cite{elitzur2019attack} utilized a CTI-based knowledge graph, based on \attack, to demonstrate increased accuracy in detecting attack patterns on enterprise networks. They used information and knowledge about past, present, and future cyber attacks that help build a comprehensive understanding of the TTPs used by cyber attackers. Gourisetti et al. \cite{gourisetti2019demonstration} developed a framework that provides functions for identifying, protecting, detecting, responding to, and recovering from cyber threats, aligning recorded events or alerts with relevant attack vectors from \attack. Gylling et al. \cite{gylling2021mapping} used \attack as the basis for their CTI when creating their probabilistic attack graph. Xiong et al.~\cite{xiong2022cyber} introduced a language to model and describe cyber threats and attacks against an enterprise security system using \attack for the enterprise. 

\textbf{Insight 6.} \attack is used to model threat scenarios and assess their impact. This helps security teams prioritize their defenses and focus on the most critical cyber risks. We believe there is a need for further research on how to effectively integrate CTI sources beyond \attack into existing security operations workflows and how to leverage the wealth of data generated by CTI for more proactive and effective threat hunting and mitigation.

\subsubsection{A-UC-PE: Product Evaluation}

Since \attack is a well-maintained knowledge base, it can be used for evaluations of cybersecurity products and research tools. Researchers have utilized \attack to evaluate security systems with scoring metrics. For example, Manocha et al.~\cite{manocha2021security} developed a security assessment rating framework that enables precise security rating for security systems. They developed a prediction score that involves weighted exploitability and impact of different levels of an attack technique. In addition, academic research has started to analyze the data stemming from \attack-based product evaluations. For example, Outkin et al.~\cite{outkin2022defender} developed a game-theoretic framework that utilizes data from MITRE's APT3 \attack Evaluations. From this data, authors were able to generalize defender capabilities.m Mashima et al.~\cite{mashima2022mitre} evaluated an in-network deception technology in a smart grid, named DecIED based on \attack for ICS. The work tests the mitigation against a few APT groups including Stuxnet and CrashOverride. It appeared that DecIED was able to mitigate only around half of the total number of \attack techniques.

\textbf{Insight 7.}  \attack is used to evaluate the capabilities of cybersecurity technologies such as assessing their ability to respond to \attack tactics and techniques. In this way, organizations can make informed decisions when choosing cybersecurity solutions. There is a lack of standardization in how cybersecurity products implement \attack and since the framework is flexible and customizable, organizations may use it differently or interpret it differently, making it challenging to compare the effectiveness of different cybersecurity products across organizations.

\vspace{-0.2cm}
\subsection{A-AD: Application Domains} \label{sec:application}
\subsubsection{A-AD-EN: Enterprise Networks}
This section discusses works that study attacks and threats related to popular CVEs of enterprise systems (vulnerabilities commonly used for internal infrastructure exploitation) \cite{ahn2020research,ampel2021linking,ajmal2021offensive}. Ahn et al.~\cite{ahn2020research} proposed a system configuration model (specific elements that define or prescribe what a system is composed of) based on the Cyber Kill Chain (CKC) and \attack to produce analytical data on threat actors resulting in providing infrastructure protection mitigation strategies. The authors utilized \attack for cyber warfare simulation and threat analysis. Xiong et al.~\cite{xiong2021cyber} proposed a threat modeling language for enterprise network security based on \attack. They analyzed key features between \attack and a Meta Attack Language framework, combining knowledge from both to define attack steps, defenses, and asset associations. Munaiah et al.~\cite{munaiah2019characterizing} carried out a penetration testing competition for enterprise systems. The authors analyzed a dataset of over $500$ million events generated by six teams of attackers during a penetration testing competition. The authors examined the competition data set to identify \attack tactics and techniques and found that it is possible to describe attackers' campaigns in a chronological sequence by analyzing their behavior.

Previous research that uses \attack has explored the connection between CVEs and \attack tactics to develop effective mitigation strategies \cite{kuppa2021linking}. Additionally, Kim et al. \cite{kim2020automated, kim2021cyber} gathered data to develop a training system for cybersecurity that focuses on threats to internal infrastructure and enterprise systems, which the simulation aims to address. The authors identified different \attack techniques and obtained scoring results for some APT groups. Hemberg et al.~\cite{hemberg2020linking} attempted to link \attack, NIST CWEs, CVEs, and CAPEC. This paper takes five browsers and compares their severity ratings, in terms of CVEs, to determine the motives behind attacks and how they will be executed. In general, most threat modeling works that use \attack, including \cite{hong2019design,outkin2021defender,bromander2020modeling,zhang2022automatic,mundt2022towards} discuss internal infrastructure attacks and the simulated threats are related to the enterprise systems' internal infrastructure. Here, Outkin et al.~\cite{outkin2021defender} developed reliable criteria for allocating resources across such detection and response opportunities at different steps in the attack. To evaluate defender policy, the authors incorporated the results of \attack Evaluation into attack success and defender response metrics.

Additionally, \attack is also being used for cloud security \cite{sahu2021model,xiong2022cyber,zhao2021exploring,zhao2022cloud,van2022identifying}. Sahu et al.~\cite{sahu2021model} developed an Infrastructure-as-a-Service (IaaS) security model named MISP, where the authors considered the \attack matrix for enterprises and a subset of it for cloud computing. The authors filtered the necessary TTPs related to the cloud for the evaluation of the adversary's behavior. Zhao et al.~\cite{zhao2021exploring,zhao2022cloud} developed a board game to improve cloud security that includes an automated evaluator to check defense plans and attack plans built by invited players. The attack cards, defense cards, and the mapping between them are derived from \attack and CSA cloud control matrix.

\subsubsection{A-AD-MCS: Mobile Communication Systems}
Mobile Communication Systems (MCS) are evolving and require standardized threat modeling frameworks~\cite{chen2021adoptability}. Authors state that \attack and Bhadra~\cite{rao2023threat} are most useful for MCS-based threat modeling. Rao et al.~\cite{rao2023threat} claimed that Bhadra aligns conveniently with \attack. Nevertheless, within the MCS domain, most works utilize \attack for threat modeling and threat detection. Early stage 5G networks have incorporated the use of \attack to demonstrate the exploitation of network functions NFV and SDN. 5G threat assessments and industry reports offer studies on how the domain-specific techniques can be used by Advanced persistent threats in multi-step attacks for 5GCN networks. Pell et al.~\cite{pell2021towards} discusses how to exploit front-facing network functions to compromise 5G networks. This work has contributed to the MITRE FiGHT, which is a knowledge base of adversary Tactics and Techniques for 5G systems \cite{mitrefight}.

\subsubsection{A-AD-ICS: Industrial Control Systems} 
Industrial Control Systems (ICS) are critical environments such as Gas, Oil, and nuclear industries. \attack literature has studied ICS and its equipment, including evaluation testbeds for ICS systems \cite{choi2020expansion,choi2021probabilistic,kwon2020cyber,gourisetti2019demonstration,att2021mitre}. Choi et al.~\cite{choi2020expansion} introduced a method to expand existing testbeds for ICS so that information can be collected during a cyber incident based on \attack. This method is useful for creating attack simulations for ICS. In a later work, authors introduced a probabilistic attack sequence generator to leverage ICS datasets \cite{choi2021probabilistic}. Here, the authors proposed a method for generating attack sequences based on the characteristics desired by the user using tactics and techniques from \attack. They overcame difficulties in developing an ICS dataset by implementing a hidden Markov model-based attack sequence generation method that uses probabilities to produce the attack sequence. Dhirani et al.~\cite{dhirani2021industrial} utilized \attack along with other standards (e.g. NIST $800-82$, ISO $27001$, IEC $62443$, etc.) to build unified an Industrial IoT standards roadmap. They specifically used \attack for identifying different aspects of ICS/SCADA security.

\vspace{-0.2cm}
\subsection{A-RF: Related Frameworks}\label{sec:correlation}
\noindent As a well-documented knowledge base of adversarial behavior, \attack has been widely adopted and combined with other cybersecurity frameworks by both academic and industrial researchers to achieve specific goals. 

\subsubsection{A-RF-CA: CAPEC} 
In addition to \attack, various threat frameworks are utilized, including the Common Attack Pattern Enumeration and Classification (CAPEC) \cite{barnum2008common}. CAPEC is a threat modeling framework that focuses on application security and is primarily associated with Common Weakness Enumeration (CWE) \cite{cwe2022}. On the other hand, \attack concentrates on network defense. Although CAPEC describes common patterns frequently employed by specific techniques described in \attack, the cross-reference helps to improve threat management by identifying potential vulnerabilities. For example, Adam et al. mapped CWEs to \attack techniques via CAPEC \cite{adam2022attack}, while Aghaei et al. \cite{aghaei2019threatzoom} created a mapping between all CVEs, CAPEC and \attack. Also, CAPEC can provide valuable insights into potential vulnerabilities within an application, while \attack can provide information on how attackers might exploit those vulnerabilities to achieve their goals.

The integration of \attack and CAPEC helps organizations to detect and mitigate a wide range of threats, including attacks against applications (which is the primary focus of CAPEC) and network infrastructure (which is the primary focus of \attack). As a result, organizations can have a more comprehensive view of the threat landscape and develop a more effective response to cyber threats. Sadlek et al. \cite{sadlek2022current} have combined CAPEC and \attack for more effective threat management. Similarly, Fox et al. \cite{fox2018enhanced} have integrated \attack and CAPEC to construct an extensive high-level threat modeling framework. Interestingly, some researchers have developed a formal knowledge base or model that unites all existing attack knowledge bases. For instance, Brazhuk et al. \cite{brazhuk2021towards} established relationships between \attack, CAPEC, CWE and CVE security enumerations to create a generic knowledge base that offers improved threat modeling over previous threat-based approaches.

\subsubsection{A-RF-CKC: Cyber Kill Chain}
\attack consists of 14 tactics that can be mapped to the phases of Lockheed Martin's Cyber Kill Chain (CKC): Reconnaissance, Weaponization, Delivery, Exploitation, Installation, Command \& Control and Actions on Objectives. Unlike the traditional CKC, \attack is a globally accessible knowledge base, which makes it more comprehensive but is also regularly updated with new techniques based on real-world observations. By understanding the different stages of an attack and the specific TTPs used by attackers, organizations can detect and prevent attacks earlier in the CKC. By mapping known TTPs to the different stages of CKC, organizations can develop a more targeted and effective response to an attack. Naik et al.~\cite{naik2022comparing} have studied characteristics, advantages and disadvantages of \attack and CKC and provide a comparative study to highlight the most suitable attack models for different applications.

\subsubsection{A-RF-ST: STRIDE}
A few works integrated multiple threat modeling frameworks for specific tasks including risk analysis and mitigation, defense framework design, and vulnerability analysis. Bolbot et al.~\cite{bolbot2022investigating} integrated \attack and STRIDE alongside cybersecurity analysis methodologies for risk analysis and mitigation. Sadlek et al.~\cite{sadlek2022identification} also used both \attack and STRIDE to identify attack paths. Straub \cite{straub2020modeling} compared the capabilities of \attack, STRIDE, and Cyber Kill Chain in the context of offensive and defensive use. He concludes that while STRIDE is useful for defensive purposes, it lacks the features required for direct offensive use. Additionally, STRIDE does not have an explicit steps to deploy an attack against the targeted vulnerability, which is a key feature of \attack. Overall, Straub's analysis suggests that while STRIDE and \attack both have their strengths and weaknesses, they serve different purposes and can be used in different ways depending on the specific goals of a given security operation.

\subsubsection{A-RF-SC: Security Controls}
To achieve threat-informed defense, native security controls can be mapped to \attack. Security Stack Mappings \cite{securitystackmappingsgithub} produces mapping files for different cloud platforms, including Microsoft Azure, Amazon Web Service, and Google Cloud Platform, to aid organizations. The online repository offers supporting resources, including scoring rubrics, mapping data formats, and mapping tools that produce the \attack navigator for mapping files. In practice, the security teams are utilizing the mapping of \attack TTPs to Azure-native
security controls~\cite{attackiq2021}. Bromander et al.~\cite{bromander2020modeling} developed a graph-based data model that linked objects obtained from \attack, STIX~\cite{barnum2012standardizing}, detection maturity model~\cite{TheDMLmo48:online} and the Diamond model~\cite{caltagirone2013diamond}. 

The National Institute of Standards and Technology (NIST) Cybersecurity Framework (CSF) was developed in 2014 and utilized to strengthen the defense and resiliency of federal networks and critical infrastructure. Kwon et al.~\cite{kwon2020cyber} proposed a Cyber Threat Dictionary that can map all attack and defense tactics to the Facility Cybersecurity Framework (FCF) through a correlation matrix~\cite{fcf2022}. FCF is specifically designed for facility-related control systems and operational technology, which motivated the authors to use the \attack for ICS matrix for mapping with FCF. 

Although \attack includes mitigation techniques against the TTPs, MITRE provides a separate and comprehensive framework named D3FEND \cite{akbar2022knowledge}, which is a knowledge graph of cybersecurity countermeasures \cite{D3FENDMa15:online}. \attack is designed from the adversaries perspective while D3FEND was built from the defenders' perspective. D3FEND was also used in academic research works such as Luh et al. \cite{luh2022penquest}.

\subsubsection{A-RF-MI: Miscellaneous}
Threat modeling frameworks include \attack, CAPEC, PASTA, WASC and OWASP. Other frameworks including Microsoft's DREAD, OCTAVE, Intel's Threat Agent Risk Assessment (TARA) and Threat Agent Library (TAL) are used for supporting security design, analysis and testing. PASTA (Process for Attack Simulation \& Threat Analysis) has been used for threat modeling in industrial IoT~\cite{wolf2021pasta}. Some frameworks including STIX (the Structured Threat Information eXpression), PRE-\attack and ODNI's CTF (Cyber Threat Framework) are useful for supporting attack information sharing. These frameworks can be integrated with \attack to generate more comprehensive and valuable risk models, thereby facilitating the identification of more effective security controls.

Jadidi et al. \cite{jadidi2021threat} proposed a unified threat-hunting model for ICS that combines
\attack for ICS and the Diamond model of intrusion analysis to predict the future behavior of the adversaries. The model can provide endpoint security logs, user behavior analytics and network or application threat analytics, which are useful to organizations. The authors evaluated their model against real-life attacks including the Ukrainian power grid attack by Black Energy 3, the DoS attack on SIEMENS PLC and Tank 101 underflow.

According to Mundt et al. \cite{mundt2022towards}, integrating CTI with Information Security Management Systems (ISMS) can result in robust data security approaches. They suggest that implementing and automating CTI processes within an ISMS can be facilitated using the \attack framework, which is commonly used by security researchers in conjunction with ISMS. To illustrate the interactions between the CTI and ISMS processes, including communication and data exfiltration, the authors use Business Process Modeling Notation (BPMN) diagrams. The proposed approach involves human actors such as a cyber analyst or Chief Information Security Officer (CISO) and follows the guidelines outlined in ISO/IEC 27000:2018. By incorporating the \attack framework into their ISMS, organizations can improve their ability to detect and respond to threats, thus enhancing their overall data security \cite{mundt2022towards}.

Hemberg et al. \cite{hemberg2020linking} proposed a framework to combine \attack, NIST, CWEs, CVEs and CAPEC. The authors proposed a bidirectional data graph named BRON to gain further insight from alerts, threats and vulnerabilities by creating links between collected information of the frameworks mentioned above. The relational links were achieved via linking \attack techniques to attack patterns, then attack patterns to CWEs and finally, CWEs to CVEs. Luh et al.~\cite{luh2022penquest} considered \attack, D3FEND and the NIST SP 800-53 to build their attack scenarios while Osquery-\attack~\cite{osqueryattackgithub} maps \attack to Osquery~\cite{osquerygithub} for enterprise threat hunting. Osquery performs continuous testing for memory leaks, thread safety and binary reproducibility on all supported platforms, including Windows, macOS and Linux (e.g. CentOS)~\cite{osqueryE95:online}. Sigma rules tagged with a $attack.tXXXX$ tag can generate the \attack Navigator~\cite{navigatorgithub} heatmap from a directory containing sigma rules. Last, the Atomic Red Team~\cite{atomicredteamgithub} is a collection of atomic tests that are mapped to \attack. The tests can be performed using command-line and aids security teams to conveniently test their environments.

\subsection{Use of \attack in Academia and Industry} 
\noindent \attack has gained significant attention from cybersecurity researchers in both academia and industry. While initially used by the industry to improve their tools and services, academic researchers have also recognized its usefulness in evaluating their research. This has resulted in a rapid development of new tools that integrate and incorporate \attack tactics and techniques. Security analysts and specialists use \attack in conjunction with other security frameworks, standards, policies, compliance and guidelines to obtain comprehensive recommendations on how to secure systems. Considering \attack as a baseline knowledge-base of TTPs, industrial research heavily involves the framework to evaluate their products, including SIEM (Security, Information, and Event Management), EDR (Endpoint Detection and Response) and deception tools. 

\emph{Cyber Threat Intelligence.} Academic researchers primarily focus on text classification and NLP for retrieving intelligence from CTI reports. 
On the other hand, the industry primarily uses \attack matrices and navigators to filter and score threats based on threat groups, techniques, platform and associated mitigation. They also develop tools and APIs (using \attack) for standard CTI sharing between organizations and developing extended threat detection tools.

\emph{Intrusion Detection}. Academia mostly attempts to categorize \attack tactics and techniques, build knowledge graphs and apply machine learning for detection and mitigation. The industry adopts and tailors \attack to develop their incident management and response tools. 

\emph{Offensive Security}. Academia uses \attack to create offensive security taxonomies, analyze past offensive security competition data and model adversarial behavior. The industry involves red and purple team exercises and organizes penetration testing using \attack. 

\emph{Cyber Risk Assessment}. Both academic and industrial researchers use \attack for assessing cyber risk by mapping threat behaviors with vulnerabilities and then proposing ways to mitigate the identified risks. Academic scholars attempt to connect other frameworks with \attack for development of better risk management. Security vendors adhere to this mapping for conducting bespoke cyber risk management for their users.

\emph{Professional Training}. Academic work focuses on theoretical analysis and modeling whereas industry addresses the training of their employees (red/purple teams), staff and clients, with practical exercises.

\emph{Threat-driven Approaches}. Academic research primarily attempts to propose new threat modeling frameworks aligned with \attack. In contrary, the industry focuses on tailoring threat modeling frameworks to use them in commercial products.

\emph{Product Evaluation}. Academia undertakes research regarding the \attack evaluation process. Industrial work involves product evaluations to determine if their developed security solutions can detect and consequently mitigate known threat actors.

\section{Research Approaches (RA) using \attack}\label{sec:res_approaches}
\noindent  In this section, we first discuss scientific methods used to build attack scenarios, models and methods based on \attack matrices. These approaches include machine learning (including natural language processing), probability theory, graph theory and game theory. Second, we study how \attack has been used in implementation of testbeds and security tools.  
Third, we study the different methods (e.g. numerical or statistical, human-based and model-based evaluations) of evaluating research that has used \attack. 

\subsection{RA-SM: Scientific Methods}\label{sec:res_methods}
\subsubsection{RA-SM-ML: Machine Learning}
Machine learning has widely been used for different \attack-based research works. Al et al.~\cite{al2020learning} used statistical machine learning analysis on APT and software attack data (270 total attack instances), reported by \attack, to identify correlations and associations among attack techniques. 
Dhir et al. \cite{dhir2021prospective} proposed to encode the labels of \attack into a set of matrices to develop the relationship between reports and labels. The authors utilized a transformer for the semantic representation of CTI reports and built causal inference to \attack. Holder et al. \cite{holder2021explainable} also focused on causal inference applied to \attack. The authors utilized explainable AI (XAI)-oriented defense recommendations and attack predictions based on \attack patterns.

ML and neural networks are often used for detection purpose. For example, Ahn et al. \cite{ahn2022malicious} performed ML-based malicious file detection and visualization based on dynamic-analysis-based \attack. Stoleriu et al. \cite{stoleriu2021cyber} proposed ML-based analysis and detection of APT attacks using ELK stack (Elasticsearch, Logstash and Kibana), where the authors retrieved a series of APT-based attacks included in the \attack matrix. Hasan et al. \cite{hasan2019artificial} developed a decision support system for cyber threat detection and protection using \attack tactics and techniques. Huang et al. \cite{huang2021open} used deep learning and \attack knowledge to develop a behavior analysis system for Windows malware. Hemberg et al. \cite{hemberg2021using} built \attack-based datasets for predicting threat techniques and attack patterns. Zurowski et al. \cite{zurowski2022quantitative} created a public dataset that includes ML-based tools, which are mapped to \attack Enterprise techniques. Bagui et al. \cite{bagui2022detecting} developed an ML-based \attack-oriented big data analysis framework for detecting reconnaissance and discovery tactics. Alnafrani et al. \cite{alnafrani2022aifis} developed an AI-based forensic investigative system, where authors used \attack to understand potential attacker capabilities. Mayami et al. \cite{maymi2017towards} created a semantic representation of adversarial TTPs, where the authors built a model of APT28 using \attack. Similarly, other works (\cite{lakhdhar2021machine,aghaei2019threatzoom}) built ML models to map vulnerabilities to adversarial tactics listed in \attack. 

\subsubsection{RA-SM-NLP: Natural Language Processing}
Natural Language Processing (NLP) is a field of study that involves the application of ML algorithms and models to analyze, understand and generate human language data. NLP has proven useful for CTI, particularly in retrieving summaries from threat reports. Liu et al.~\cite{liu2022threat} used an attention transformer hierarchical recurrent neural network to extract \attack information from CTI. Kuppa et al.~\cite{kuppa2021linking} employed NLP techniques, such as the Multi-Head Joint Embedding Neural Network model, to automatically map CVEs to \attack techniques. Chen et al. \cite{chen2022building} developed an anomaly detection and threat hunting system that utilizes NLP and graph modeling. The authors used \attack APT3 evaluation data and applied NLP techniques to process Windows logs for seeking suspicious patterns. Niakanlahiji et al. \cite{niakanlahiji2018natural} presented an NLP-based trend analysis to present how to obtain knowledge regarding APTs from unstructured reports and developed an information retrieval system named SECCMiner that combines NLP processes and information retrieval system concepts to categorize APTs based on \attack tactics. Husari et al. \cite{husari2019learning} also utilized NLP to characterize the temporal relationship of attack actions of an APT using \attack and a machine readable language named STIX. Apart from the above-mentioned works, \cite{ayoade2018automated,purba2020word,karuna2021automating,otgonpurev2021effective,sadlek2022current,chen2022building,domschot2022automated} have involved NLP and \attack for automated threat intelligence, modeling and mapping.

\subsubsection{RA-SM-PT: Probability Theory}

Choi et al.~\cite{choi2021probabilistic} utilized a hidden Markov model to generate varied attack sequences based on user objectives. The authors considered the probability of starting each \attack tactic as the initial state probability, probability of movement between each tactic as transition probability and probability of the occurrence of a particular technique (under the same tactics) as the emission probability. The attack sequence generation can leverage ICS datasets and provide various attack scenarios performed in real life by different malware including Stuxnet (Iran nuclear facilities), BlackEnergy3 \& Industroyer (Ukraine power grid), Triton (Saudi Arabia petrochemical plant), Bad Rabbit (Ukrainian transportation) and LockerGoga (Norway aluminum company). Other works including \cite{ahmed2022mitre,evensjo2020probability,he2021model} calculated probabilities of different attack scenarios to assess and mitigate risks. All these works recognized \attack as a standard knowledge base of TTPs and utilized listed tactics and techniques for their simulated attack scenarios.

\subsubsection{RA-SM-GM: Graph Modeling}
Kriaa et al.~\cite{kriaa2021seckg} used graph theory due to the complex nature of APTs and comprehensive attack methods, which provides a better evaluation than some other existing methods. Here, the authors combined knowledge graphs and machine learning to detect and prevent adversarial techniques. Xiong et al.~\cite{xiong2021cyber} proposed algorithms and graph-based mapping to provide insights into certain attacks, such as MAL file Access Token Manipulation. The authors proposed a threat modeling language called enterpriseLang, which presents a domain-specific language based on the Meta Attack Language (MAL~\cite{johnson2018meta}) framework. Here, MAL is directly associated with \attack and leverages TTPs to define attack steps in the language. Hacks et al.~\cite{hacks2021integrating} proposed an approach for integrating user actions and security behavior to attack simulations by mapping Security Behavior Analysis (SBA) to MAL through \attack techniques. Hemberg et al.~\cite{hemberg2020linking} proposed a graph-based linking technique called BRON that links \attack techniques to attack patterns, patterns to weaknesses and weaknesses to CVEs. 
More works, including \cite{fairbanks2021att,fairbanks2021identifying,dravsar2020session} utilized \attack and graphs for threat intelligence and modeling.

\subsubsection{RA-SM-GT: Game Theory}
Outkin et al. \cite{outkin2022defender} proposed a game-theoretic framework, called GPLADD, to constantly allocate resources (e.g. to sensing and assessment of attack indicators) against an uncertain stream of attacks. The attack data used for the evaluation of the framework are from \attack. 
Nisioti et al.~\cite{nisioti2021game} utilized game theory to determine optimal investigating policies after a cyber incident. The proposed framework considers the cost for investigating an \attack technique and available actions for the investigator with the attacker type and anti-forensics techniques being unknown. 
Luh et al. \cite{luh2022penquest} proposed a game-theoretic framework, called PenQuest, to support security education and cyber risk assessment by simulating a game whether an attacker attempts to compromise an infrastructure and the defender attempts to protect it. As in the previous papers, attack data for the evaluation of PenQuest were drawn from \attack.

\vspace{-0.25cm}
\subsection{RA-I: Implementations}\label{sec:implementation}
\subsubsection{RA-I-TE: Testbeds}
Choi et al.~\cite{choi2020expansion} outlined vulnerabilities and threats for ICS and implemented a testbed to help fix cybersecurity issues by offering a better understanding of how to mitigate vulnerabilities. The authors set 52 techniques excluding duplicates in ten tactics mapped to 92 intrusion detection rules using the \attack for Enterprise. Hong et al.~\cite{hong2019design} implemented a testbed where a simulated threat generator automatically generates cyber threats based on \attack to help improve the coping ability of system security officers in dealing with cyber threats. Here, the threat generator allows for the addition of evolving cyber threats and the selection of the next threat. Halverson et al.~\cite{halvorsen2019evaluating} developed a testbed to evaluate the effectiveness of their developed tool TOMATO, which uses MITRE \attack to simulate attacks and evaluate the observability and efficiency of a set of deployed monitoring techniques. The approach was integrated into an ELK stack, and evaluated on real SCADA devices within the Washington State University smart city testbed. Most papers that use \attack for offensive security or professional training implemented a testbed to deploy the attack scenarios. For example, Ajmal et al.~\cite{ajmal2021offensive} developed a simulated environment to implement different attack scenarios. Luh et al.~\cite{luh2022penquest} developed a testbed for human experimentation-based evaluation of their proposed game model. Drašar et al.~\cite{dravsar2020session} created a small-scale network to emulate various \attack-based attack scenarios.

\subsubsection{RA-I-TO: Tools}
\attack aids adversarial emulation and consequent defensive tools that can assess certain attack scenarios. \attack has been used to design and develop certain adversary emulation tools including Red or Purple team tools. Defensive tools are also designed and developed considering \attack-based tactics and techniques. Halvorsen et al.~\cite{halvorsen2019evaluating} developed the TOMATO (Threat Observability and Monitoring Assessment) tool that can evaluate the observability of network security monitoring strategies. TOMATO provides observability scores and monitoring technique efficiency scores while using \attack-based simulated attacks.

\emph{Red Teaming Tools}. There are a few open-source \attack test tools including CALDERA \cite{calderagithub}, Endgame Red teaming Automation \cite{rtagithub}, Red Canary Atomic Red \cite{atomicredteamgithub} and Uber Metta \cite{mettagithub}. These tools have adapted \attack and provided platforms for red teams to simulate attacks. Each tool features a different set of tactics for penetrating a network and helps the administrator find out the security weaknesses or entry points. Since \attack itself is always under development, these tools follow the same path, and new features are added on a regular basis.

\emph{Purple Teaming Tools}. Purple Team \attack Automation \cite{purpleteamattackgithub} is another automated adversary tactics emulation platform that is built on top of the Metasploit framework \cite{metasploitgithub}. The platform integrated codes and techniques from \attack, tools like CALDERA, and libraries like the Atomic Red Team \cite{atomicredteamgithub} and RE:TERNAL \cite{reternalgithub}, which is a centralized purple team orchestration service to test blue-team capabilities against red-team techniques. All included simulations of the tool are mapped and aligned to \attack. There are other \attack-oriented tools as well, which are used for generating detection rules (e.g. sigma rules). For example, S2AN \cite{s2angithub} is a standalone tool that creates an \attack Navigator \cite{navigatorgithub} based on a directory containing sigma rules \cite{sigmagithub} and Suricata signatures. Kriaa et al.~\cite{kriaa2021seckg} used the Grakn tool to create targeted knowledge graphs and query them using the \textit{graql} language. The authors built a knowledge graph for their proposed approach using \attack, to gather knowledge on attacks from different sources. This offers capabilities to detect attack techniques and then learn to predict them by processing event logs. 

Appropriate datasets are necessary to aid the community with mapping real data to open source projects such as Sigma, Atomic Red Team, Threat Hunter Playbook, and \attack knowledge base. The project entitled \textit{Security Datasets} \cite{securitydatasetsgithub} is an open-source dataset collection that facilitates adversary emulation, enables security and threat actor analysis and adversarial behavior, and provides datasets for Capture-The-Flag (CTF) competitions.

\vspace{-0.2cm}
\subsection{RA-E: Evaluations}\label{sec:evaluation}
\subsubsection{RA-E-NE: Numeric Evaluation}
Al et al.~\cite{al2020learning} utilized hierarchical clustering to investigate the association among techniques included in \attack and later discovered 98 different clusters representing these associations. The authors evaluated the mutual information (of the techniques in the fine-grain clusters, as well as the coarse-grain clusters directly from the datasets) by measuring fine-grain associations (within the same cluster) for APTs using both technique-based and cluster-based normalized mutual information (NMI). The \textit{maximum predictability} of each technique can be calculated based on its cluster assignment. 

Hemberg et al.~\cite{hemberg2020linking} evaluated their graph model through different statistical analyses. The number of edges (links) is calculated as they connect different \attack techniques, patterns, weaknesses, and associated CVEs. By calculating the query times for threats connected to the Top 10 CVEs, threats, and vulnerabilities for the top 25 CWEs and riskiest software, the authors measured the relational linkage statistics for tactics, techniques, and attack patterns over the number of edges in the graph. Similarly, the authors measured the counts and distributions of vulnerability connections and affected product configurations.

Kim et al.~\cite{kim2021cyber} provided a severity scoring methodology for APT-based and fileless cyber attacks and later evaluated the scores with the cyber kill chain and \attack. The authors evaluated APTs and fileless cyberattacks that occurred between 2010 and 2020. They calculated scores for the APT groups: powerliks, Rozena, Duqu 2.0, Kovter, Petya, Sorebrect, WannaCry, Magniber, Emotet and Gandcrab. 

\subsubsection{RA-E-HE: Human Evaluation}
Hacks et al.~\cite{hacks2021integrating} integrated human behavior analysis to the attack simulations and attempted to calculate probabilities of an attack being successful. Authors conducted surveys where employees would answer the questionnaire of Security Behavior Analysis and their answers were given as inputs to a vulnerability assessment tool for conducting attack simulation on an IT infrastructure. Further, evaluating a model or association by domain experts is often helpful. Al et al.~\cite{al2020learning} recruited six domain experts with at least five years of experience and knowledge in the area of cyber threat intelligence and \attack. According to the experts, 93\% of the fine-grain associations of \attack techniques (within the same cluster) and 90\% of the coarse-grain associations (inter-cluster) present strong correlations, which validates their way of utilizing hierarchical clustering techniques. 

Oconnor \cite{oconnor2022helo} developed a lab (e.g. post-exploit lab) for practicing and improving experiential learning, payloads writing, detection evasion, attack functionality, post-exploitation tools development and network traffic manipulation based on \attack. The author discussed ethical issues and introduced to the students the Computer and Fraud Abuse Act (CFAA), Electronic Communications Privacy Act (ECPA), the Digital Millennium Copyright Act (DCMA) and the corresponding university's acceptable use policy. Last, Luh et al. \cite{luh2022penquest} involved students to evaluate their proposed game theoretic model for technical education.

\subsubsection{RA-E-ME: Model Evaluation}
Most of the works that use \attack to develop a threat model, later evaluate it based on the reliability in providing security assessments and suggesting security settings. For example, Xiong et al.~\cite{xiong2021cyber} evaluated enterpriseLang by modeling two attack scenarios: the Ukraine cyber attack of 2015 and the Cayman National Bank cyber heist of 2016. The authors used the Enterprise \attack matrix as a knowledge base for the proposed language. 

Choi et al.~\cite{choi2021probabilistic} evaluated their Hidden Markov Model-based attack sequence generator by validating whether the attack sequence from initial access to impact follows the pattern of real-life malware. The authors adopted \attack to design the attack sequence. They confirmed that this model generated the actual attack sequence of Triton, which was discovered in the Saudi Arabia petrochemical plant. Ampel et al.~\cite{ampel2021linking} compared their CVET model against benchmark classical machine learning, deep learning, and pre-trained language models for text classification tasks to understand how these models perform while linking CVEs to \attack. The authors showed that CVET achieves the highest accuracy (76.93\%) and F1-score (76.18\%) among the compared models.

 \section{Conclusions and Future Directions}\label{sec:lessons}
\noindent This paper provides a comprehensive review of research and industry applications of the MITRE \attack framework and proposes a taxonomy for categorizing literature that uses \attack. 
In this section, we will recapitulate the essential points discussed in the preceding sections and delve into contemporary challenges, constraints and potential future research works associated with \attack.

\emph{Holistic Approach}. \attack takes a holistic approach to cybersecurity, covering defensive and offensive techniques. It provides a comprehensive list of adversary tactics and techniques used in cyber attacks, making it an essential resource for threat intelligence, threat modeling, risk assessment and offensive security. This approach makes it possible to understand better the adversary and their tactics, which is essential for developing effective defense strategies.
    
\emph{Open and Community-Driven}. \attack is continually updated based on community feedback and contributions. This collaborative approach ensures that the framework remains up-to-date and relevant, providing organizations with the latest information on adversary tactics and techniques.
    
\emph{Common Medium for Knowledge Sharing}. \attack provides a common language for the cybersecurity industry in terms of threat intelligence, making it easier for organizations to communicate and collaborate on cybersecurity. This common language ensures that everyone is on the same page, which is essential for effective communication and collaboration in the emerging threat landscape.
    
\emph{Wide Coverage}. \attack covers a wide range of attack techniques across different platforms and technologies, including Windows, Linux and macOS. It also covers ICS, Cloud and Mobile platforms. This broad coverage makes it a valuable resource for organizations with different IT environments.
    
\emph{Mapping to Other Frameworks}. As we have seen in the existing literature, \attack can be mapped to other cybersecurity frameworks, such as the NIST Cybersecurity Framework, ISO/IEC 27001, COBIT, etc. This mapping provides a way to merge different frameworks to achieve particular needs.
    
\emph{Flexibility}. \attack is customizable. It enables organizations to utilize it to their specific needs such as create custom intelligence, threat models, risk assessments and offensive security strategies. This customization makes the framework more relevant and valuable to specific organizations and industries.

Even though \attack is a reputed knowledge base of TTPs, there are a few limitations of it. 

\emph{Evolving Threat Landscape}. The threat landscape constantly evolves and attackers are constantly developing new TTPs. \attack may only sometimes reflect the latest threats and must be updated regularly to stay current.
    
\emph{Limited Geographical Coverage}. \attack is based on observations of attacks that have taken place in the United States, Europe and other developed regions. The tactics and techniques used by attackers in other parts of the world may need to be better represented in the framework. 
    
\emph{Focus on Specific Threat Actors}. \attack focuses on a limited set of well-known threat actors and may need to fully capture the tactics and techniques used by other, less well-known groups.
    
\emph{Tactical Level}. \attack provides a tactical-level view of adversary tactics and techniques and does not provide a comprehensive view of the overall attack lifecycle.

Despite the widespread adoption of \attack for improved threat mitigation and prevention, there remain a few untapped scenarios for researchers and developers to contribute to. The following require answers in future studies.

\emph{Real-time Threat Intelligence and Incident Response}. Real-time threat intelligence is critical for quickly detecting and responding to attacks. We find a lack of research that utilizes \attack to address these issues. Researchers and experts can work on developing real-time threat intelligence capabilities that can leverage \attack to identify and respond to attacks more quickly.

\emph{Risk Quantification}. Cyber risk quantification (CRQ) is a major challenge in both fields of cyber research and in the progress of industry. Accurate risk assessments are crucial for ensuring effective spending, as demonstrated by the demands of Chief Information Security Officers (CISOs). Extended research on \attack can address this challenge by incorporating CRQ methods that are derived from \attack and seamlessly integrate threat behaviors and quantitative data from threat intelligence sources, such as threat event frequency \cite{freund2014measuring}.

\emph{Collaboration between Academic and Industrial Research}. \attack provides a platform for academic and industry stakeholders to showcase the performance of their methods or implement their software and serve their clients. However, ongoing revision and expansion of the framework concepts and data are essential to keep up with evolving threat behaviors. Integrating industry and academic perspectives through collaboration is crucial in determining the need for new techniques or tactics and developing effective mitigation methods against newly discovered threat mechanisms or sub-techniques. One way, researchers can reduce the gap is to develop new ideas in academia and evaluate these new frameworks, techniques, or workflows in the industry.

\emph{Developing Industry-specific Threat Models}. While \attack covers a wide range of platforms and technologies, industry-specific threat models can provide a more tailored approach to identifying and responding to attacks in particular attack scenarios. Researchers and experts can further work on developing industry-specific threat models that leverage \attack.

With the evolution and improvement of language models and chatbots like chatGPT, cyber threat intelligence and modeling have new areas to explore. Overall, emerging technologies lead to continuously new cyber threats and \attack is required to be updated on a regular basis. Each of these updates can initiate new research directions for academic researchers and industrial experts.

We believe that future research and development will benefit from close collaborations between academic and industrial researchers. For example, academia can utilize theoretical attack and defense models that involve \attack and then the industry can test their products against these models. Likewise, the industry can share data, from the evaluation of their products, with academic scholars for fostering novel scientific ideas in the field, which can then feed back to their products and services.

\vspace{-0.2cm}
\bibliographystyle{IEEEtran}
\bibliography{references}

\begin{thebibliography}{100}
\providecommand{\url}[1]{#1}
\csname url@samestyle\endcsname
\providecommand{\newblock}{\relax}
\providecommand{\bibinfo}[2]{#2}
\providecommand{\BIBentrySTDinterwordspacing}{\spaceskip=0pt\relax}
\providecommand{\BIBentryALTinterwordstretchfactor}{4}
\providecommand{\BIBentryALTinterwordspacing}{\spaceskip=\fontdimen2\font plus
\BIBentryALTinterwordstretchfactor\fontdimen3\font minus
  \fontdimen4\font\relax}
\providecommand{\BIBforeignlanguage}[2]{{%
\expandafter\ifx\csname l@#1\endcsname\relax
\typeout{** WARNING: IEEEtran.bst: No hyphenation pattern has been}%
\typeout{** loaded for the language `#1'. Using the pattern for}%
\typeout{** the default language instead.}%
\else
\language=\csname l@#1\endcsname
\fi
#2}}
\providecommand{\BIBdecl}{\relax}
\BIBdecl

\bibitem{georgiadou2021assessing}
A.~Georgiadou, S.~Mouzakitis, and D.~Askounis, ``Assessing {MITRE} {ATT\&CK}
  risk using a cyber-security culture framework,'' \emph{Sensors}, vol.~21,
  no.~9, p. 3267, 2021.

\bibitem{bodeau2018cyber}
D.~J. Bodeau, C.~D. McCollum, and D.~B. Fox, ``Cyber threat modeling: Survey,
  assessment, and representative framework,'' MITRE CORP MCLEAN VA MCLEAN,
  Tech. Rep., 2018.

\bibitem{sans2022}
M.~Bromiley, ``{SANS} 2022 {ATT\&CK} and {D3FEND} report: Incorporating
  frameworks into your analysis and intelligence,'' {SANS} Institute, January
  2022.

\bibitem{choi2020expansion}
S.~Choi, J.~Choi, J.-H. Yun, B.-G. Min, and H.~Kim, ``Expansion of {ICS}
  testbed for security validation based on {MITRE} {ATT\&CK} techniques,'' in
  \emph{13th USENIX Workshop on Cyber Security Experimentation and Test
  (CSET)}, 2020.

\bibitem{al2020learning}
R.~Al-Shaer, J.~M. Spring, and E.~Christou, ``Learning the associations of
  {MITRE} {ATT\&CK} adversarial techniques,'' in \emph{2020 IEEE Conference on
  Communications and Network Security (CNS)}.\hskip 1em plus 0.5em minus
  0.4em\relax IEEE, 2020, pp. 1--9.

\bibitem{pennington2019getting}
A.~Pennington, A.~Applebaum, K.~Nickels, T.~Schulz, B.~Strom, and J.~Wunder,
  ``Getting started with {ATT\&CK},'' MITRE Corp, McLean, VA, Tech. Rep., 2019.

\bibitem{alexander2020mitre}
O.~Alexander, M.~Belisle, and J.~Steele, ``{MITRE} {ATT\&CK} for industrial
  control systems: Design and philosophy,'' \emph{MITRE Corporation, Bedford,
  MA, USA}, 2020.

\bibitem{claroty2021}
CLAROTY, ``Supporting the {MITRE ATT\&CK} for {ICS} framework,''
  \url{https://security.claroty.com/white-paper/supporting-mitre-ics}, 2021.

\bibitem{cylance2019}
Cylance, ``How to use the {MITRE} {ATT\&CK} enterprise framework,'' {Research
  Desk},
  \url{https://www.demandtalk.com/whitepaper/it-infra/how-to-use-the-mitre-attck-enterprise-framework/},
  October 2019.

\bibitem{cisco2021}
C.~Secure, ``Why endpoint security is critical to today’s ciso,'' {Cisco
  Public},
  \url{https://www.cisco.com/c/en/us/products/collateral/security/white-paper-c11-744950.pdf},
  May 2021.

\bibitem{fortinet2021}
FORTINET, ``Assess your endpoint security,''
  \url{https://www.fortinet.com/content/dam/fortinet/assets/white-papers/wp-assess-your-endoint-security.pdf},
  March 2022.

\bibitem{attackiq2021}
ATTACKIQ, ``Leveraging {MITRE} {ATT\&CK} to secure the cloud,''
  \url{https://attackiq.com/lp/leveraging-mitre-attack-to-secure-the-cloud/},
  2021.

\bibitem{parmar2019use}
M.~Parmar and A.~Domingo, ``On the use of cyber threat intelligence (cti) in
  support of developing the commander's understanding of the adversary,'' in
  \emph{MILCOM 2019-2019 IEEE Military Communications Conference
  (MILCOM)}.\hskip 1em plus 0.5em minus 0.4em\relax IEEE, 2019, pp. 1--6.

\bibitem{fox2018enhanced}
D.~B. Fox, E.~I. Arnoth, C.~W. Skorupka, C.~D. McCollum, and D.~Bodeau,
  ``Enhanced cyber threat model for financial services sector ({FSS})
  institutions,'' \emph{The Homeland Security Systems Engineering and
  Development Institute, McLean, VA, USA}, 2018.

\bibitem{wagner2019cyber}
T.~D. Wagner, K.~Mahbub, E.~Palomar, and A.~E. Abdallah, ``Cyber threat
  intelligence sharing: Survey and research directions,'' \emph{Computers \&
  Security}, vol.~87, p. 101589, 2019.

\bibitem{schlette2021comparative}
D.~Schlette, M.~Caselli, and G.~Pernul, ``A comparative study on cyber threat
  intelligence: The security incident response perspective,'' \emph{IEEE
  Communications Surveys \& Tutorials}, vol.~23, no.~4, pp. 2525--2556, 2021.

\bibitem{cascavilla2021cybercrime}
G.~Cascavilla, D.~A. Tamburri, and W.-J. Van Den~Heuvel, ``Cybercrime threat
  intelligence: A systematic multi-vocal literature review,'' \emph{Computers
  \& Security}, vol. 105, p. 102258, 2021.

\bibitem{ibrahim2020challenges}
A.~Ibrahim, D.~Thiruvady, J.~G. Schneider, and M.~Abdelrazek, ``The challenges
  of leveraging threat intelligence to stop data breaches,'' \emph{Frontiers in
  Computer Science}, vol.~2, p.~36, 2020.

\bibitem{dutta2020overview}
A.~Dutta and S.~Kant, ``An overview of cyber threat intelligence platform and
  role of artificial intelligence and machine learning,'' in
  \emph{International Conference on Information Systems Security}.\hskip 1em
  plus 0.5em minus 0.4em\relax Springer, 2020, pp. 81--86.

\bibitem{zibak2022threat}
A.~Zibak, C.~Sauerwein, and A.~C. Simpson, ``Threat intelligence quality
  dimensions for research and practice,'' \emph{Digital Threats: Research and
  Practice}, 2022.

\bibitem{abu2018cyber}
M.~S. Abu, S.~R. Selamat, A.~Ariffin, and R.~Yusof, ``Cyber threat
  intelligence--issue and challenges,'' \emph{Indonesian Journal of Electrical
  Engineering and Computer Science}, vol.~10, no.~1, pp. 371--379, 2018.

\bibitem{tounsi2018survey}
W.~Tounsi and H.~Rais, ``A survey on technical threat intelligence in the age
  of sophisticated cyber attacks,'' \emph{Computers \& security}, vol.~72, pp.
  212--233, 2018.

\bibitem{mavroeidis2017cyber}
V.~Mavroeidis and S.~Bromander, ``Cyber threat intelligence model: an
  evaluation of taxonomies, sharing standards, and ontologies within cyber
  threat intelligence,'' in \emph{2017 European Intelligence and Security
  Informatics Conference (EISIC)}.\hskip 1em plus 0.5em minus 0.4em\relax IEEE,
  2017, pp. 91--98.

\bibitem{brown20212021}
R.~Brown and R.~M. Lee, ``2021 {SANS} cyber threat intelligence survey,'' in
  \emph{Tech. Rep}.\hskip 1em plus 0.5em minus 0.4em\relax SANS Institute,
  2021.

\bibitem{brown2019evolution}
------, ``The evolution of cyber threat intelligence ({CTI}): 2019 {SANS} {CTI}
  survey,'' \emph{SANS Institute. Available online:
  \url{https://www.sans.org/white-papers/38790/}, Accessed on July 12, 2021.},
  2019.

\bibitem{shackleford2017cyber}
D.~Shackleford, ``Cyber threat intelligence uses, successes and failures: The
  {SANS} 2017 {CTI} survey,'' \emph{SANS Institute}, 2017.

\bibitem{tayouri2023survey}
D.~Tayouri, N.~Baum, A.~Shabtai, and R.~Puzis, ``A survey of mulval extensions
  and their attack scenarios coverage,'' \emph{IEEE Access}, 2023.

\bibitem{sadlek2022current}
L.~Sadlek, P.~{\v{C}}eleda, and D.~Tovar{\v{n}}{\'a}k, ``Current challenges of
  cyber threat and vulnerability identification using public enumerations,'' in
  \emph{Proceedings of the 17th International Conference on Availability,
  Reliability and Security}, 2022, pp. 1--8.

\bibitem{ahn2020research}
M.~K. Ahn and J.~R. Lee, ``Research on system architecture and methodology
  based on {MITRE} {ATT\&CK} for experiment analysis on cyber warfare
  simulation,'' \emph{Journal of the Korea Society of Computer and
  Information}, vol.~25, no.~8, pp. 31--37, 2020.

\bibitem{ampel2021linking}
B.~Ampel, S.~Samtani, S.~Ullman, and H.~Chen, ``Linking common vulnerabilities
  and exposures to the {MITRE} {ATT\&CK} framework: A self-distillation
  approach,'' \emph{arXiv preprint arXiv:2108.01696}, 2021.

\bibitem{hong2019design}
S.~Hong, K.~Kim, and T.~Kim, ``The design and implementation of simulated
  threat generator based on {MITRE} {ATT\&CK} for cyber warfare training,''
  \emph{Journal of the Korea Institute of Military Science and Technology},
  vol.~22, no.~6, pp. 797--805, 2019.

\bibitem{kuppa2021linking}
A.~Kuppa, L.~Aouad, and N.-A. Le-Khac, ``Linking {CVE}'s to {MITRE} {ATT\&CK}
  techniques,'' in \emph{The 16th International Conference on Availability,
  Reliability and Security}, 2021, pp. 1--12.

\bibitem{munaiah2019characterizing}
N.~Munaiah, A.~Rahman, J.~Pelletier, L.~Williams, and A.~Meneely,
  ``Characterizing attacker behavior in a cybersecurity penetration testing
  competition,'' in \emph{2019 ACM/IEEE International Symposium on Empirical
  Software Engineering and Measurement (ESEM)}.\hskip 1em plus 0.5em minus
  0.4em\relax IEEE, 2019, pp. 1--6.

\bibitem{outkin2021defender}
A.~V. Outkin, P.~V. Schulz, T.~Schulz, T.~D. Tarman, and A.~Pinar, ``Defender
  policy evaluation and resource allocation using {MITRE} {ATT\&CK} evaluations
  data,'' \emph{arXiv preprint arXiv:2107.04075}, 2021.

\bibitem{pell2021towards}
R.~Pell, S.~Moschoyiannis, E.~Panaousis, and R.~Heartfield, ``Towards dynamic
  threat modelling in 5g core networks based on {MITRE} att\&ck,'' \emph{arXiv
  preprint arXiv:2108.11206}, 2021.

\bibitem{xiong2021cyber}
W.~Xiong, E.~Legrand, O.~{\AA}berg, and R.~Lagerstr{\"o}m, ``Cyber security
  threat modeling based on the {MITRE} enterprise {ATT\&CK} matrix,''
  \emph{Software and Systems Modeling}, pp. 1--21, 2021.

\bibitem{hemberg2020linking}
E.~Hemberg, J.~Kelly, M.~Shlapentokh-Rothman, B.~Reinstadler, K.~Xu, N.~Rutar,
  and U.-M. O'Reilly, ``Linking threat tactics, techniques, and patterns with
  defensive weaknesses, vulnerabilities and affected platform configurations
  for cyber hunting,'' \emph{arXiv preprint arXiv:2010.00533}, 2020.

\bibitem{kim2021cyber}
K.~Kim, F.~A. Alfouzan, and H.~Kim, ``Cyber-attack scoring model based on the
  offensive cybersecurity framework,'' \emph{Applied Sciences}, vol.~11,
  no.~16, p. 7738, 2021.

\bibitem{choi2021probabilistic}
S.~Choi, J.-H. Yun, and B.-G. Min, ``Probabilistic attack sequence generation
  and execution based on {MITRE} {ATT\&CK} for ics datasets,'' in \emph{Cyber
  Security Experimentation and Test Workshop}, 2021, pp. 41--48.

\bibitem{arshad2021attack}
S.~Arshad, M.~Alam, S.~Al-Kuwari, and M.~H.~A. Khan, ``Attack specification
  language: Domain specific language for dynamic training in cyber range,'' in
  \emph{2021 IEEE Global Engineering Education Conference (EDUCON)}.\hskip 1em
  plus 0.5em minus 0.4em\relax IEEE, 2021, pp. 873--879.

\bibitem{golushko2020application}
A.~P. Golushko and V.~G. Zhukov, ``Application of advanced persistent threat
  actorstechniques aor evaluating defensive countermeasures,'' in \emph{2020
  IEEE Conference of Russian Young Researchers in Electrical and Electronic
  Engineering (EIConRus)}.\hskip 1em plus 0.5em minus 0.4em\relax IEEE, 2020,
  pp. 312--317.

\bibitem{kriaa2021seckg}
S.~Kriaa and Y.~Chaabane, ``Sec{KG}: Leveraging attack detection and prediction
  using knowledge graphs,'' in \emph{2021 12th International Conference on
  Information and Communication Systems (ICICS)}.\hskip 1em plus 0.5em minus
  0.4em\relax IEEE, 2021, pp. 112--119.

\bibitem{kwon2020cyber}
R.~Kwon, T.~Ashley, J.~Castleberry, P.~Mckenzie, and S.~N.~G. Gourisetti,
  ``Cyber threat dictionary using {MITRE} {ATT\&CK} matrix and {NIST}
  cybersecurity framework mapping,'' in \emph{2020 Resilience Week
  (RWS)}.\hskip 1em plus 0.5em minus 0.4em\relax IEEE, 2020, pp. 106--112.

\bibitem{bromander2020modeling}
S.~Bromander, M.~Swimmer, M.~Eian, G.~Skjotskift, and F.~Borg, ``Modeling cyber
  threat intelligence.'' in \emph{ICISSP}, 2020, pp. 273--280.

\bibitem{legoy2020automated}
V.~Legoy, M.~Caselli, C.~Seifert, and A.~Peter, ``Automated retrieval of
  {ATT\&CK} tactics and techniques for cyber threat reports,'' \emph{arXiv
  preprint arXiv:2004.14322}, 2020.

\bibitem{fairbanks2021att}
J.~Fairbanks, A.~Orbe, C.~Patterson, E.~Serra, and M.~Scheepers, ``Att\&ck
  tactics in android malware control flow graph through graph representation
  learning and interpretability.'' in \emph{Proceedings of the 2021 IEEE
  International Conference on Big Data (REU 2021 Symposium)}, 2021.

\bibitem{huang2021open}
Y.-T. Huang, C.~Y. Lin, Y.-R. Guo, K.-C. Lo, Y.~S. Sun, and M.~C. Chen, ``Open
  source intelligence for malicious behavior discovery and interpretation,''
  \emph{IEEE Transactions on Dependable and Secure Computing}, 2021.

\bibitem{kurniawan2021att}
K.~Kurniawan, A.~Ekelhart, and E.~Kiesling, ``An att\&ck-kg for linking
  cybersecurity attacks to adversary tactics and techniques,'' 2021.

\bibitem{lakhdhar2021machine}
Y.~Lakhdhar and S.~Rekhis, ``Machine learning based approach for the automated
  mapping of discovered vulnerabilities to adversial tactics,'' in \emph{2021
  IEEE Security and Privacy Workshops (SPW)}.\hskip 1em plus 0.5em minus
  0.4em\relax IEEE, 2021, pp. 309--317.

\bibitem{lee2021fileless}
G.~Lee, S.~Shim, B.~Cho, T.~Kim, and K.~Kim, ``Fileless cyberattacks: Analysis
  and classification,'' \emph{ETRI Journal}, vol.~43, no.~2, pp. 332--343,
  2021.

\bibitem{mendsaikhan2020automatic}
O.~Mendsaikhan, H.~Hasegawa, Y.~Yamaguchi, and H.~Shimada, ``Automatic mapping
  of vulnerability information to adversary techniques,'' in \emph{The
  Fourteenth International Conference on Emerging Security Information, Systems
  and Technologies SECUREWARE2020}, 2020.

\bibitem{purba2020word}
M.~D. Purba, B.~Chu, and E.~Al-Shaer, ``From word embedding to cyber-phrase
  embedding: Comparison of processing cybersecurity texts,'' in \emph{2020 IEEE
  International Conference on Intelligence and Security Informatics
  (ISI)}.\hskip 1em plus 0.5em minus 0.4em\relax IEEE, 2020, pp. 1--6.

\bibitem{aghaei2019threatzoom}
E.~Aghaei and E.~Al-Shaer, ``Threatzoom: neural network for automated
  vulnerability mitigation,'' in \emph{Proceedings of the 6th Annual Symposium
  on Hot Topics in the Science of Security}, 2019, pp. 1--3.

\bibitem{ajmal2021offensive}
A.~B. Ajmal, M.~A. Shah, C.~Maple, M.~N. Asghar, and S.~U. Islam, ``Offensive
  security: Towards proactive threat hunting via adversary emulation,''
  \emph{IEEE Access}, vol.~9, pp. 126\,023--126\,033, 2021.

\bibitem{brazhuk2021towards}
A.~Brazhuk, ``Towards automation of threat modeling based on a semantic model
  of attack patterns and weaknesses,'' \emph{arXiv preprint arXiv:2112.04231},
  2021.

\bibitem{elitzur2019attack}
A.~Elitzur, R.~Puzis, and P.~Zilberman, ``Attack hypothesis generation,'' in
  \emph{2019 European Intelligence and Security Informatics Conference
  (EISIC)}.\hskip 1em plus 0.5em minus 0.4em\relax IEEE, 2019, pp. 40--47.

\bibitem{fairbanks2021identifying}
J.~Fairbanks, A.~Orbe, C.~Patterson, J.~Layne, E.~Serra, and M.~Scheepers,
  ``Identifying {ATT\&CK} tactics in android malware control flow graph through
  graph representation learning and interpretability,'' in \emph{2021 IEEE
  International Conference on Big Data (Big Data)}.\hskip 1em plus 0.5em minus
  0.4em\relax IEEE, 2021, pp. 5602--5608.

\bibitem{franklin2017toward}
L.~Franklin, M.~Pirrung, L.~Blaha, M.~Dowling, and M.~Feng, ``Toward a
  visualization-supported workflow for cyber alert management using threat
  models and human-centered design,'' in \emph{2017 IEEE Symposium on
  Visualization for Cyber Security (VizSec)}.\hskip 1em plus 0.5em minus
  0.4em\relax IEEE, 2017, pp. 1--8.

\bibitem{gourisetti2019demonstration}
S.~N.~G. Gourisetti, M.~Mylrea, T.~Ashley, R.~Kwon, J.~Castleberry,
  Q.~Wright-Mockler, P.~McKenzie, and G.~Brege, ``Demonstration of the
  cybersecurity framework through real-world cyber attack,'' in \emph{2019
  Resilience Week (RWS)}, vol.~1.\hskip 1em plus 0.5em minus 0.4em\relax IEEE,
  2019, pp. 19--25.

\bibitem{gylling2021mapping}
A.~Gylling, M.~Ekstedt, Z.~Afzal, and P.~Eliasson, ``Mapping cyber threat
  intelligence to probabilistic attack graphs,'' in \emph{2021 IEEE
  International Conference on Cyber Security and Resilience (CSR)}.\hskip 1em
  plus 0.5em minus 0.4em\relax IEEE, 2021, pp. 304--311.

\bibitem{hacks2021integrating}
S.~Hacks, I.~Butun, R.~Lagerstr{\"o}m, A.~Buhaiu, A.~Georgiadou, and
  A.~Michalitsi~Psarrou, ``Integrating security behavior into attack
  simulations,'' in \emph{The 16th International Conference on Availability,
  Reliability and Security}, 2021, pp. 1--13.

\bibitem{hacks2022measuring}
S.~Hacks, L.~Persson, and N.~Hers{\'e}n, ``Measuring and achieving test
  coverage of attack simulations extended version,'' \emph{Software and Systems
  Modeling}, pp. 1--16, 2022.

\bibitem{hassanzadeh2018samiit}
A.~Hassanzadeh and R.~Burkett, ``{SAMIIT}: Spiral attack model in {IIoT}
  mapping security alerts to attack life cycle phases,'' in \emph{5th
  International Symposium for ICS \& SCADA Cyber Security Research 2018 5},
  2018, pp. 11--20.

\bibitem{ahmed2022mitre}
M.~Ahmed, S.~Panda, C.~Xenakis, and E.~Panaousis, ``{MITRE} {ATT\&CK}-driven
  cyber risk assessment,'' in \emph{Proceedings of the 17th International
  Conference on Availability, Reliability and Security}, 2022, pp. 1--10.

\bibitem{bolbot2022investigating}
V.~Bolbot, S.~Basnet, H.~Zhao, O.~V. Banda, and B.~Silverajan, ``Investigating
  a novel approach for cybersecurity risk analysis with application to remote
  pilotage operations,'' in \emph{European Workshop on Maritime Systems
  Resilience and Security}, 2022.

\bibitem{oruc2022assessing}
A.~Oruc, A.~Amro, and V.~Gkioulos, ``Assessing cyber risks of an {INS} using
  the {MITRE} {ATT\&CK} framework,'' \emph{Sensors}, vol.~22, no.~22, p. 8745,
  2022.

\bibitem{oconnor2022helo}
T.~OConnor, ``Helo darkside: Breaking free from katas and embracing the
  adversarial mindset in cybersecurity education,'' in \emph{Proceedings of the
  53rd ACM Technical Symposium on Computer Science Education V. 1}, 2022, pp.
  710--716.

\bibitem{kim2020automated}
D.~Kim, Y.~Kim, M.-K. Ahn, and H.~Lee, ``Automated cyber threat emulation based
  on {ATT\&CK} for cyber security training,'' \emph{Journal of the Korea
  Society of Computer and Information}, vol.~25, no.~9, pp. 71--80, 2020.

\bibitem{rao2023threat}
S.~P. Rao, H.-Y. Chen, and T.~Aura, ``Threat modeling framework for mobile
  communication systems,'' \emph{Computers \& Security}, vol. 125, p. 103047,
  2023.

\bibitem{chen2022building}
C.~K. Chen, S.~C. Lin, S.~C. Huang, Y.~T. Chu, C.~L. Lei, and C.~Y. Huang,
  ``Building machine learning-based threat hunting system from scratch,''
  \emph{Digital Threats: Research and Practice}, 2022.

\bibitem{adam2022attack}
C.~Adam, M.~F. Bulut, D.~Sow, S.~Ocepek, C.~Bedell, and L.~Ngweta, ``Attack
  techniques and threat identification for vulnerabilities,'' \emph{arXiv
  preprint arXiv:2206.11171}, 2022.

\bibitem{sadlek2022identification}
L.~Sadlek, P.~{\v{C}}eleda, and D.~Tovar{\v{n}}{\'a}k, ``Identification of
  attack paths using kill chain and attack graphs,'' in \emph{2022-2022
  IEEE/IFIP Network Operations and Management Symposium (NOMS)}.\hskip 1em plus
  0.5em minus 0.4em\relax IEEE, 2022, pp. 1--6.

\bibitem{jadidi2021threat}
Z.~Jadidi and Y.~Lu, ``A threat hunting framework for industrial control
  systems,'' \emph{IEEE Access}, vol.~9, pp. 164\,118--164\,130, 2021.

\bibitem{mundt2022towards}
M.~Mundt and H.~Baier, ``Towards mitigation of data exfiltration techniques
  using the {MITRE} {ATT\&CK} framework,'' in \emph{International Conference on
  Digital Forensics and Cyber Crime}.\hskip 1em plus 0.5em minus 0.4em\relax
  Springer, 2022, pp. 139--158.

\bibitem{niakanlahiji2018natural}
A.~Niakanlahiji, J.~Wei, and B.-T. Chu, ``A natural language processing based
  trend analysis of advanced persistent threat techniques,'' in \emph{2018 IEEE
  International Conference on Big Data (Big Data)}.\hskip 1em plus 0.5em minus
  0.4em\relax IEEE, 2018, pp. 2995--3000.

\bibitem{ayoade2018automated}
G.~Ayoade, S.~Chandra, L.~Khan, K.~Hamlen, and B.~Thuraisingham, ``Automated
  threat report classification over multi-source data,'' in \emph{2018 IEEE 4th
  International Conference on Collaboration and Internet Computing
  (CIC)}.\hskip 1em plus 0.5em minus 0.4em\relax IEEE, 2018, pp. 236--245.

\bibitem{karuna2021automating}
P.~Karuna, E.~Hemberg, U.~M. O'Reilly, and N.~Rutar, ``Automating cyber threat
  hunting using {NLP}, automated query generation, and genetic perturbation,''
  \emph{arXiv preprint arXiv:2104.11576}, 2021.

\bibitem{shin2021art}
Y.~Shin, K.~Kim, J.~J. Lee, and K.~Lee, ``Art: Automated reclassification for
  threat actors based on {ATT\&CK} matrix similarity,'' in \emph{2021 World
  Automation Congress (WAC)}.\hskip 1em plus 0.5em minus 0.4em\relax IEEE,
  2021, pp. 15--20.

\bibitem{he2021model}
T.~He and Z.~Li, ``A model and method of information system security risk
  assessment based on {MITRE} {ATT\&CK},'' in \emph{2021 2nd International
  Conference on Electronics, Communications and Information Technology
  (CECIT)}.\hskip 1em plus 0.5em minus 0.4em\relax IEEE, 2021, pp. 81--86.

\bibitem{johnson2018meta}
P.~Johnson, R.~Lagerstr{\"o}m, and M.~Ekstedt, ``A meta language for threat
  modeling and attack simulations,'' in \emph{Proceedings of the 13th
  International Conference on Availability, Reliability and Security}, 2018,
  pp. 1--8.

\bibitem{manocha2021security}
H.~Manocha, A.~Srivastava, C.~Verma, R.~Gupta, and B.~Bansal, ``Security
  assessment rating framework for enterprises using {MITRE} {ATT\&CK} matrix,''
  \emph{arXiv preprint arXiv:2108.06559}, 2021.

\bibitem{mashima2022mitre}
D.~Mashima, ``{MITRE} {ATT\&CK} based evaluation on in-network deception
  technology for modernized electrical substation systems,''
  \emph{Sustainability}, vol.~14, no.~3, p. 1256, 2022.

\bibitem{dhirani2021industrial}
L.~L. Dhirani, E.~Armstrong, and T.~Newe, ``Industrial iot, cyber threats, and
  standards landscape: evaluation and roadmap,'' \emph{Sensors}, vol.~21,
  no.~11, p. 3901, 2021.

\bibitem{luh2022penquest}
R.~Luh, S.~Eresheim, S.~Gr{\"o}{\ss}bacher, T.~Petelin, F.~Mayr, P.~Tavolato,
  and S.~Schrittwieser, ``{PenQuest} reloaded: A digital cyber defense game for
  technical education,'' in \emph{2022 IEEE Global Engineering Education
  Conference (EDUCON)}.\hskip 1em plus 0.5em minus 0.4em\relax IEEE, 2022, pp.
  906--914.

\bibitem{husari2019learning}
G.~Husari, E.~Al-Shaer, B.~Chu, and R.~F. Rahman, ``Learning {APT} chains from
  cyber threat intelligence,'' in \emph{Proceedings of the 6th Annual Symposium
  on Hot Topics in the Science of Security}, 2019, pp. 1--2.

\bibitem{nisioti2021game}
A.~Nisioti, G.~Loukas, S.~Rass, and E.~Panaousis, ``Game-theoretic decision
  support for cyber forensic investigations,'' \emph{Sensors}, vol.~21, no.~16,
  p. 5300, 2021.

\bibitem{halvorsen2019evaluating}
J.~Halvorsen, J.~Waite, and A.~Hahn, ``Evaluating the observability of network
  security monitoring strategies with {TOMATO},'' \emph{IEEE Access}, vol.~7,
  pp. 108\,304--108\,315, 2019.

\bibitem{wong2021threat}
A.~Y. Wong, E.~G. Chekole, M.~Ochoa, and J.~Zhou, ``Threat modeling and
  security analysis of containers: A survey,'' \emph{arXiv preprint
  arXiv:2111.11475}, 2021.

\bibitem{dhir2021prospective}
N.~Dhir, H.~Hoeltgebaum, N.~Adams, M.~Briers, A.~Burke, and P.~Jones,
  ``Prospective artificial intelligence approaches for active cyber defence,''
  \emph{arXiv preprint arXiv:2104.09981}, 2021.

\bibitem{holder2021explainable}
E.~Holder and N.~Wang, ``Explainable artificial intelligence ({XAI})
  interactively working with humans as a junior cyber analyst,''
  \emph{Human-Intelligent Systems Integration}, vol.~3, no.~2, pp. 139--153,
  2021.

\bibitem{ahn2022malicious}
G.~Ahn, K.~Kim, W.~Park, and D.~Shin, ``Malicious file detection method using
  machine learning and interworking with {MITRE} {ATT\&CK} framework,''
  \emph{Applied Sciences}, vol.~12, no.~21, p. 10761, 2022.

\bibitem{stoleriu2021cyber}
R.~Stoleriu, A.~Puncioiu, and I.~Bica, ``Cyber attacks detection using open
  source {ELK} stack,'' in \emph{2021 13th International Conference on
  Electronics, Computers and Artificial Intelligence (ECAI)}.\hskip 1em plus
  0.5em minus 0.4em\relax IEEE, 2021, pp. 1--6.

\bibitem{bagui2022detecting}
S.~Bagui, D.~Mink, S.~Bagui, T.~Ghosh, T.~McElroy, E.~Paredes, N.~Khasnavis,
  and R.~Plenkers, ``Detecting reconnaissance and discovery tactics from the
  {MITRE} {ATT\&CK} framework in {Zeek} {Conn} {Logs} using {Spark}'s machine
  learning in the big data framework,'' \emph{Sensors}, vol.~22, no.~20, p.
  7999, 2022.

\bibitem{zurowski2022quantitative}
S.~Zurowski, G.~Lord, and I.~Baggili, ``A quantitative analysis of offensive
  cyber operation ({OCO}) automation tools,'' in \emph{Proceedings of the 17th
  International Conference on Availability, Reliability and Security}, 2022,
  pp. 1--11.

\bibitem{alnafrani2022aifis}
R.~Alnafrani and D.~Wijesekera, ``{AIFIS}: Artificial intelligence ({AI})-based
  forensic investigative system,'' in \emph{2022 10th International Symposium
  on Digital Forensics and Security (ISDFS)}.\hskip 1em plus 0.5em minus
  0.4em\relax IEEE, 2022, pp. 1--6.

\bibitem{samtani2022explainable}
S.~Samtani, H.~Chen, M.~Kantarcioglu, and B.~Thuraisingham, ``Explainable
  artificial intelligence for cyber threat intelligence ({XAI}-{CTI}),''
  \emph{IEEE Transactions on Dependable and Secure Computing}, vol.~19, no.~04,
  pp. 2149--2150, 2022.

\bibitem{grigorescu2022cve2att}
O.~Grigorescu, A.~Nica, M.~Dascalu, and R.~Rughinis, ``{CVE2ATT\&CK}:
  {BERT}-based mapping of {CVEs} to {MITRE} {ATT\&CK} techniques,''
  \emph{Algorithms}, vol.~15, no.~9, p. 314, 2022.

\bibitem{hasan2019artificial}
K.~Hasan, S.~Shetty, and S.~Ullah, ``Artificial intelligence empowered cyber
  threat detection and protection for power utilities,'' in \emph{2019 IEEE 5th
  International Conference on Collaboration and Internet Computing
  (CIC)}.\hskip 1em plus 0.5em minus 0.4em\relax IEEE, 2019, pp. 354--359.

\bibitem{maymi2017towards}
F.~Maym{\'\i}, R.~Bixler, R.~Jones, and S.~Lathrop, ``Towards a definition of
  cyberspace tactics, techniques and procedures,'' in \emph{2017 IEEE
  International Conference on Big Data (Big Data)}.\hskip 1em plus 0.5em minus
  0.4em\relax IEEE, 2017, pp. 4674--4679.

\bibitem{dravsar2020session}
M.~Dra{\v{s}}ar, S.~Moskal, S.~Yang, and P.~Zat'ko, ``Session-level adversary
  intent-driven cyberattack simulator,'' in \emph{2020 IEEE/ACM 24th
  International Symposium on Distributed Simulation and Real Time Applications
  (DS-RT)}.\hskip 1em plus 0.5em minus 0.4em\relax IEEE, 2020, pp. 1--9.

\bibitem{kim2022comparative}
H.~Kim, H.~Kim \emph{et~al.}, ``Comparative experiment on {TTP} classification
  with class imbalance using oversampling from {CTI} dataset,'' \emph{Security
  and Communication Networks}, vol. 2022, 2022.

\bibitem{kim2021automatically}
K.~Kim, Y.~Shin, J.~Lee, and K.~Lee, ``Automatically attributing mobile threat
  actors by vectorized {ATT\&CK} matrix and paired indicator,'' \emph{Sensors},
  vol.~21, no.~19, p. 6522, 2021.

\bibitem{sahu2021model}
I.~K. Sahu and M.~J. Nene, ``Model for {IaaS} security model: {MISP}
  framework,'' in \emph{2021 International Conference on Intelligent
  Technologies (CONIT)}.\hskip 1em plus 0.5em minus 0.4em\relax IEEE, 2021, pp.
  1--6.

\bibitem{zhao2021exploring}
T.~Zhao, T.~E. Gasiba, U.~Lechner, and M.~Pinto-Albuquerque, ``Exploring a
  board game to improve cloud security training in industry (short paper),'' in
  \emph{Second International Computer Programming Education Conference (ICPEC
  2021)}.\hskip 1em plus 0.5em minus 0.4em\relax Schloss
  Dagstuhl-Leibniz-Zentrum f{\"u}r Informatik, 2021.

\bibitem{zhao2022cloud}
T.~Zhao, U.~Lechner, M.~Pinto-Albuquerque, and E.~Ata, ``Cloud of assets and
  threats: A playful method to raise awareness for cloud security in
  industry,'' \emph{OpenAccess Series in Informatics}, 2022.

\bibitem{van2022identifying}
G.~van~der Merwe, C.~Muller, W.~van~der Merwe, and D.~Blaauw, ``Identifying
  adversaries' signatures using knowledge representations of cyberattack
  techniques on cloud infrastructure,'' in \emph{International Conference on
  Cyber Warfare and Security}, vol.~17, no.~1, 2022, pp. 333--339.

\bibitem{zhang2022automatic}
S.~Zhang, P.~Chen, G.~Bai, S.~Wang, M.~Zhang, S.~Li, and C.~Zhao, ``An
  automatic assessment method of cyber threat intelligence combined with
  {ATT\&CK} matrix,'' \emph{Wireless Communications and Mobile Computing}, vol.
  2022, 2022.

\bibitem{odemis2022detecting}
M.~Odemis, C.~Yucel, and A.~Koltuksuz, ``Detecting user behavior in cyber
  threat intelligence: development of {Honeypsy} system,'' \emph{Security and
  Communication Networks}, vol. 2022, 2022.

\bibitem{jo2022cyberattack}
Y.~Jo, O.~Choi, J.~You, Y.~Cha, and D.~H. Lee, ``Cyberattack models for ship
  equipment based on the {MITRE} {ATT\&CK} framework,'' \emph{Sensors},
  vol.~22, no.~5, p. 1860, 2022.

\bibitem{dettectgithub}
R.~C.~D. Centre, ``{DeTTECT},'' \url{https://github.com/rabobank-cdc/DeTTECT},
  2022, (Accessed on 16/12/2022).

\bibitem{DeTTCTMa95:online}
``{DeTT\&CT}: Mapping detection to {MITRE} {ATT\&CK},'' NVISO Labs,
  \url{https://blog.nviso.eu/2022/03/09/dettct-mapping-detection-to-mitre-attck/},
  March 2022, (Accessed on 09/21/2022).

\bibitem{roberts2021structured}
A.~Roberts, ``Structured intelligence--what does it even mean?'' in \emph{Cyber
  Threat Intelligence}.\hskip 1em plus 0.5em minus 0.4em\relax Springer, 2021,
  pp. 37--64.

\bibitem{farooq2018optimal}
H.~M. Farooq and N.~M. Otaibi, ``Optimal machine learning algorithms for cyber
  threat detection,'' in \emph{2018 UKSim-AMSS 20th International Conference on
  Computer Modelling and Simulation (UKSim)}.\hskip 1em plus 0.5em minus
  0.4em\relax IEEE, 2018, pp. 32--37.

\bibitem{HowMITRE75:online}
A.~R. Sharma, ``How {MITRE} {ATT\&CK} alignment supercharges your {SIEM},''
  Securonix,
  \url{https://www.securonix.com/blog/how-mitre-attck-alignment-supercharges-your-siem},
  (Accessed on 10/19/2022).

\bibitem{sadrazamis2022mitre}
K.~Sadrazamis, ``{MITRE} {ATT\&CK}-based analysis of cyber-attacks in
  intelligent transportation,'' 2022.

\bibitem{amro2023assessing}
A.~Amro, V.~Gkioulos, and S.~Katsikas, ``Assessing cyber risk in cyber-physical
  systems using the att\&ck framework,'' \emph{ACM Transactions on Privacy and
  Security}, vol.~26, no.~2, pp. 1--33, 2023.

\bibitem{kure2022integrated}
H.~I. Kure, S.~Islam, and H.~Mouratidis, ``An integrated cyber security risk
  management framework and risk predication for the critical infrastructure
  protection,'' \emph{Neural Computing and Applications}, pp. 1--31, 2022.

\bibitem{RiskandV25:online}
TLP.White, ``Risk and vulnerability assessment ({RVA}) mapped to the {MITRE}
  {ATT\&CK} framework infographic,''
  \url{https://www.cisa.gov/sites/default/files/publications/FY19_RVAs_Mapped_to_the_MITRE_ATTCK_Framework_508.pdf},
  (Accessed on 11/06/2022).

\bibitem{RVAsMapp12:online}
------, ``{RVAs} mapped to the {MITRE} {ATT\&CK} framework,''
  \url{https://irp.cdn-website.com/9a5fc83f/files/uploaded/FY20_RVAs_Mapped_to_the_MITRE_ATTCK_Framework_508_QVzrjj9OT2e6JWUkrOAu.pdf},
  (Accessed on 11/06/2022).

\bibitem{grantek12:online}
{GRANTEK}, ``{RVAs} mapped to the {MITRE} {ATT\&CK} framework,''
  \url{https://grantek.com/wp-content/uploads/2020/04/2020CybersecurityWP.pdf},
  April 2020.

\bibitem{ciso12:online}
{AttackIQ}, ``The {CISO}'s guide to better vulnerability management using
  {MITRE} {ATT\&CK},''
  \url{https://www.attackiq.com/wp-content/uploads/2021/12/90398r72vt8w.pdf},
  December 2021.

\bibitem{xiong2022cyber}
W.~Xiong, E.~Legrand, O.~{\AA}berg, and R.~Lagerstr{\"o}m, ``Cyber security
  threat modeling based on the {MITRE} {Enterprise} {ATT\&CK} matrix,''
  \emph{Software and Systems Modeling}, vol.~21, no.~1, pp. 157--177, 2022.

\bibitem{outkin2022defender}
A.~V. Outkin, P.~V. Schulz, T.~Schulz, T.~D. Tarman, and A.~Pinar, ``Defender
  policy evaluation and resource allocation with {MITRE} {ATT\&CK} evaluations
  data,'' \emph{IEEE Transactions on Dependable and Secure Computing}, 2022.

\bibitem{chen2021adoptability}
H.~Y. Chen and S.~P. Rao, ``On adoptability and use case exploration of threat
  modeling for mobile communication systems,'' in \emph{Proceedings of the 2021
  ACM SIGSAC Conference on Computer and Communications Security}, 2021, pp.
  2417--2419.

\bibitem{mitrefight}
{MITRE Corporation}, ``{FiGHT} ({5G} hierarchy of threats),''
  \url{https://fight.mitre.org}, 2022, (Online; accessed 16-December-2022).

\bibitem{att2021mitre}
------, ``{MITRE} {ATT\&CK},'' \url{https://attack. mitre. org}, 2015-2022,
  (Online; accessed 16-December-2022).

\bibitem{barnum2008common}
M.~S. Barnum, ``Common attack pattern enumeration and classification (capec)
  schema,'' \emph{Department of Homeland Security}, 2008.

\bibitem{cwe2022}
{MITRE Corporation}, ``{Common} {Weakness} {Enumeration},''
  \url{https://cwe.mitre.org}, 2022, (Accessed on 16/12/2022).

\bibitem{naik2022comparing}
N.~Naik, P.~Jenkins, P.~Grace, and J.~Song, ``Comparing attack models for it
  systems: Lockheed martin’s cyber kill chain, mitre att\&ck framework and
  diamond model,'' in \emph{2022 IEEE International Symposium on Systems
  Engineering (ISSE)}.\hskip 1em plus 0.5em minus 0.4em\relax IEEE, 2022, pp.
  1--7.

\bibitem{straub2020modeling}
J.~Straub, ``Modeling attack, defense and threat trees and the cyber kill
  chain, att\&ck and stride frameworks as blackboard architecture networks,''
  in \emph{2020 IEEE International Conference on Smart Cloud
  (SmartCloud)}.\hskip 1em plus 0.5em minus 0.4em\relax IEEE, 2020, pp.
  148--153.

\bibitem{securitystackmappingsgithub}
T.~C. for Threat-Informed~Defense, ``Security stack mappings,''
  \url{https://github.com/center-for-threat-informed-defense/security-stack-mappings},
  2022, (Accessed on 16/12/2022).

\bibitem{barnum2012standardizing}
S.~Barnum, ``Standardizing cyber threat intelligence information with the
  structured threat information expression ({STIX}),'' \emph{MITRE
  Corporation}, vol.~11, pp. 1--22, 2012.

\bibitem{TheDMLmo48:online}
R.~Stillions, ``The {DML} model,''
  \url{http://ryanstillions.blogspot.com/2014/04/the-dml-model_21.html},
  (Accessed on 10/05/2022).

\bibitem{caltagirone2013diamond}
S.~Caltagirone, A.~Pendergast, and C.~Betz, ``The diamond model of intrusion
  analysis,'' Center For Cyber Intelligence Analysis and Threat Research,
  Hanover, MD, Tech. Rep., 2013.

\bibitem{fcf2022}
F.~Cybersecurity, ``Facility cybersecurity framework,''
  \url{https://facilitycyber.labworks.org}, 2022, (Accessed on 16/12/2022).

\bibitem{akbar2022knowledge}
K.~A. Akbar, S.~M. Halim, Y.~Hu, A.~Singhal, L.~Khan, and B.~Thuraisingham,
  ``Knowledge mining in cybersecurity: From attack to defense,'' in \emph{IFIP
  Annual Conference on Data and Applications Security and Privacy}.\hskip 1em
  plus 0.5em minus 0.4em\relax Springer, 2022, pp. 110--122.

\bibitem{D3FENDMa15:online}
``D3fend matrix | {MITRE} d3fend™,'' \url{https://d3fend.mitre.org/},
  (Accessed on 07/21/2022).

\bibitem{wolf2021pasta}
A.~Wolf, D.~Simopoulos, L.~D'Avino, and P.~Schwaiger, ``The pasta threat model
  implementation in the iot development life cycle,'' \emph{INFORMATIK 2020},
  2021.

\bibitem{osqueryattackgithub}
F.~Mottini, ``{Osquery}-{ATT\&CK},''
  \url{https://github.com/teoseller/osquery-attck}, 2022, (Accessed on
  16/12/2022).

\bibitem{osquerygithub}
Osquery, ``{Osquery},'' \url{https://github.com/osquery/osquery}, 2022,
  (Accessed on 16/12/2022).

\bibitem{osqueryE95:online}
``osquery: Easily ask questions about your {Linux}, {Windows}, and {macOS}
  infrastructure,'' \url{https://osquery.io/}, (Accessed on 09/21/2022).

\bibitem{navigatorgithub}
{MITRE Corporation}, ``{MITRE} {ATT\&CK} navigator,''
  \url{https://mitre-attack.github.io/attack-navigator/}, 2021, (Accessed on
  16/12/2022).

\bibitem{atomicredteamgithub}
{Red Canary}, ``Atomic red team,''
  \url{https://github.com/redcanaryco/atomic-red-team}, 2019, [Online; accessed
  16-December-2022].

\bibitem{hemberg2021using}
E.~Hemberg and U.-M. O'Reilly, ``Using a collated cybersecurity dataset for
  machine learning and artificial intelligence,'' \emph{arXiv preprint
  arXiv:2108.02618}, 2021.

\bibitem{liu2022threat}
C.~Liu, J.~Wang, and X.~Chen, ``Threat intelligence {ATT\&CK} extraction based
  on the attention transformer hierarchical recurrent neural network,''
  \emph{Applied Soft Computing}, vol. 122, p. 108826, 2022.

\bibitem{otgonpurev2021effective}
M.~Otgonpurev, ``Effective application of natural language processing
  techniques in automated cyber threat intelligence,'' 2021.

\bibitem{domschot2022automated}
E.~Domschot, ``Automated labeling of {MITRE} {ATT\&CK} tactics and techniques
  in malware threat reports,'' Ph.D. dissertation, New Mexico Institute of
  Mining and Technology, 2022.

\bibitem{evensjo2020probability}
L.~Evensj{\"o}, ``Probability analysis and financial model development of
  {MITRE} {ATT\&CK} enterprise matrix's attack steps and mitigations,'' 2020.

\bibitem{calderagithub}
{MITRE Corporation}, ``Caldera,'' \url{https://github.com/mitre/caldera}, 2022,
  (Accessed on 16/12/2022).

\bibitem{rtagithub}
Endgame, ``Red team automation,'' \url{https://github.com/endgameinc/RTA},
  2022, (Accessed on 16/12/2022).

\bibitem{mettagithub}
U.~Common, ``Metta,'' \url{https://github.com/uber-common/metta}, 2018,
  [Online; accessed 16-December-2022].

\bibitem{purpleteamattackgithub}
Praetorian, ``Purple team {ATT\&CK™} automation,''
  \url{https://github.com/praetorian-inc/purple-team-attack-automation}, 2020,
  (Online; accessed 16-December-2022).

\bibitem{metasploitgithub}
Rapid7, ``{The Metasploit Framework},''
  \url{https://github.com/rapid7/metasploit-framework}, 2020, (Accessed on
  16/12/2022).

\bibitem{reternalgithub}
J.~Dreijer, ``{RE:TERNAL},''
  \url{https://github.com/d3vzer0/reternal-quickstart}, 2020, (Accessed on
  16/12/2022).

\bibitem{s2angithub}
3CORESec, ``{S2AN},'' \url{https://github.com/3CORESec/S2AN}, 2021, [Online;
  accessed 16-December-2022].

\bibitem{sigmagithub}
SigmaHQ, ``{Sigma},'' \url{https://github.com/SigmaHQ/sigma}, 2022, (Accessed
  on 16/12/2022).

\bibitem{securitydatasetsgithub}
O.~T.~R. Forge, ``{Security Datasets},''
  \url{https://github.com/OTRF/Security-Datasets}, 2022, (Accessed on
  16/12/2022).

\bibitem{freund2014measuring}
J.~Freund and J.~Jones, \emph{Measuring and managing information risk: a FAIR
  approach}.\hskip 1em plus 0.5em minus 0.4em\relax Butterworth-Heinemann,
  2014.

\end{thebibliography}
\end{document}